\documentclass[preprint,journal]{vgtc}             
\date{February 2026}
\manuscriptnote{}
\usepackage{amsmath} 
\usepackage{enumitem}
\usepackage{amsfonts}
\usepackage[utf8]{inputenc}
\usepackage[T1]{fontenc}
\usepackage{xspace}
\usepackage{array}
\usepackage{booktabs}   
\usepackage{multirow}   
\usepackage{caption}    
\usepackage{geometry}   

\onlineid{1927}

\preprinttext{}

\usepackage[scaled=0.92]{inconsolata}
\usepackage{minted}
\setminted{
  fontsize=\footnotesize,
  breaklines=true,
  breakanywhere=true,
  autogobble=true,
  linenos=false,
  frame=none,
  xleftmargin=0.75em,
  baselinestretch=0.95
}

\newcommand{\ourrep}{Flint\xspace}

\vgtccategory{Research}

\title{\texorpdfstring{\ourrep: A Semantics-Driven Data Visualization \\Intermediate Language\\
{\large\mdseries\href{https://microsoft.github.io/flint-chart/}{microsoft.github.io/flint-chart/}}}
{Flint: A Semantics-Driven Data Visualization Intermediate Language}}

\author{%
   \authororcid{Yunhai Wang}{0000-0003-0059-6580},
   \authororcid{Kecheng Lu}{0000-0001-5990-3296},
    \authororcid{Junhao Chen}{0009-0002-3177-0385},
   \authororcid{Alper Sarikaya}{0000-0002-2876-4286},
\authororcid{Chenglong Wang}{0000-0002-5933-6620}
}

\authorfooter{
  \item Yunhai Wang, Kecheng Lu, and Junhao Chen are with Renmin University of China, Beijing, China. E-mail: \{wang.yh, lukecheng0407, chenjunhao11\}@ruc.edu.cn.
  \item Alper Sarikaya and Chenglong Wang are with Microsoft, Redmond, WA, USA. E-mail: \{alper.sarikaya, chenglong.wang\}@microsoft.com.
}

\abstract{%
We present \ourrep, an intermediate language that enables authors to create high-quality visualizations from concise, semantics-driven specifications without explicitly configuring low-level parameters such as scales, axes, and formatting. Unlike prior systems that infer default configurations from surface-level data representations, often producing brittle choices, \ourrep introduces a hierarchical data semantic model that allows users to specify the meanings of data fields structurally and helps the compiler derive appropriate visualization configurations. From a concise specification, the system generates and optimizes library-agnostic visualization configurations and translates them into complete, executable specifications for multiple target grammars, including Vega-Lite, Apache ECharts, and Chart.js. We demonstrate that \ourrep simplifies the authoring process without compromising on visual quality, and it is an effective intermediate language for both humans and AI agents to create visualizations.

}


\teaser{
  \centering
  \includegraphics[width=\linewidth]{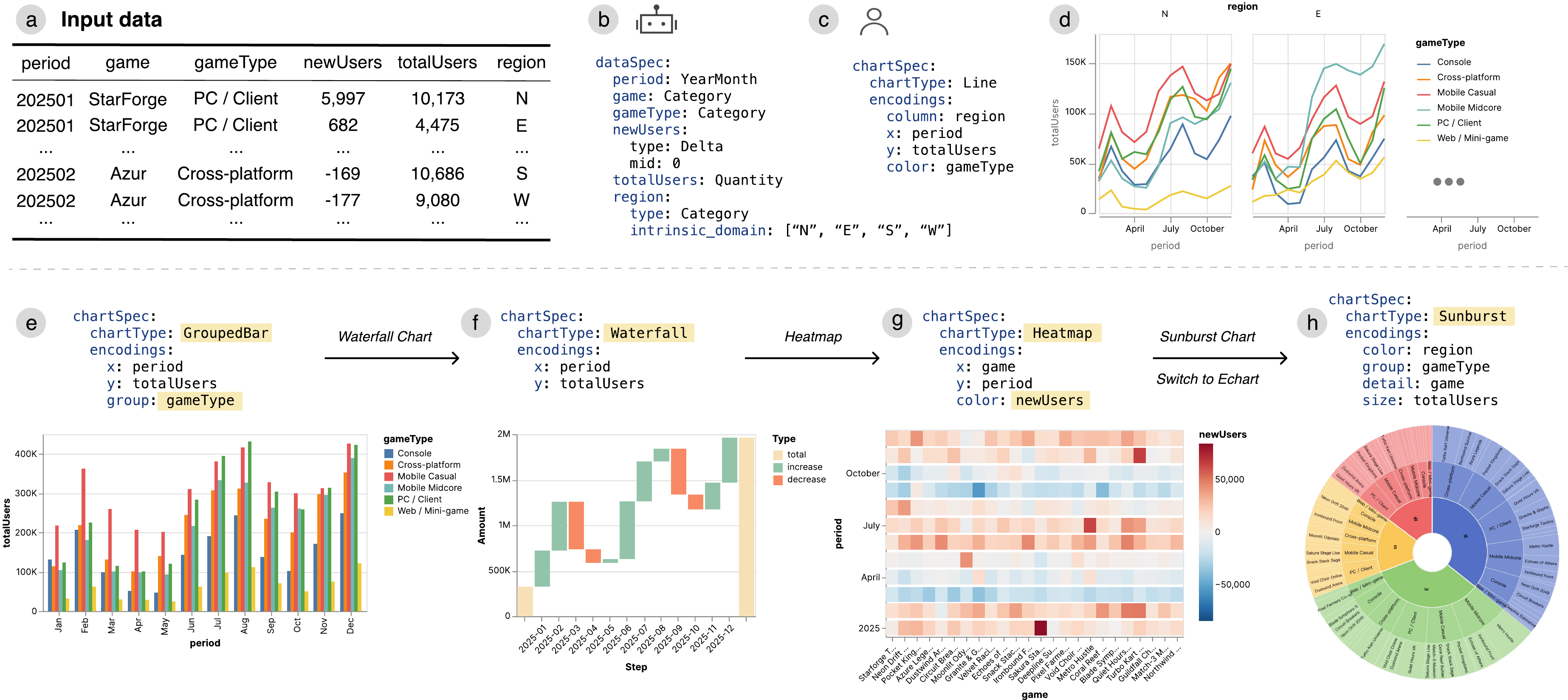}\vspace{-2mm}
    \caption{\ourrep is an intermediate language for data visualization. A \ourrep specification has two parts: a data spec, which can be inferred by an LLM agent from the data, and a simple chart spec. Given the input data, the compiler uses semantic information to choose low-level parameters and translate the \ourrep spec into a well-configured 
    \look{specification for a target chart type, such as a Vega-Lite faceted line chart.}
    As the user edits the \ourrep spec to create new visualizations, the compiler automatically updates these configurations to produce \emph{high-quality charts} without requiring the user to manually adjust detailed parameters such as axes, layers, or color schemes. This allows users to rapidly explore data across diverse chart types and rendering backends, for example by switching to ECharts to support a sunburst chart.}    \label{fig:case-study}
}

\graphicspath{{figs/}{figures/}{pictures/}{images/}{./}} 

\usepackage{tcolorbox}

\usepackage{mathptmx}                  

\newcommand{\bpstart}[1]{\smallskip\noindent{\textbf{#1.}}}

\newcommand{\look}[1]{{{#1}}}

\newcommand{\code}[1]{{\footnotesize \textsf{#1}}}
\begin{document}



\firstsection{Introduction}
\maketitle
Effective data exploration requires analysts to rapidly create and adapt visualizations, frequently changing chart types, encodings, and layouts as they refine hypotheses. Consequently, visualization specifications must remain simple while maintaining the capacity to produce high-quality designs.
Declarative languages like Vega-Lite~\cite{satyanarayan2016vega} and ECharts~\cite{li2018echarts} have advanced this goal, with recent work combining these grammars with large language models (LLMs) to enable natural language to visualization (NL2Vis) systems~\cite{brossier2026state} and autonomous data agents~\cite{vaithilingam2024dynavis,wang2025data}. These approaches lower the barrier to iterative chart creation by generating complete specifications from user descriptions.

Despite these advances, authors still face a persistent trade-off. Simple high-level specifications that fully rely on system ``smart default'' low-level configurations~\cite{wilkinson2006grammar} often yield poor designs, whereas detailed specifications can improve quality but become verbose, brittle, and difficult to modify because of complex parameter dependencies. The fundamental issue is that existing libraries must infer data semantics from \emph{raw representations} to configure smart defaults. Incorrect inferences produce misleading visualizations: stacking non-additive measures like correlation, applying a sequential color scale to diverging-class fields, or parsing integer dates (e.g., 20080101) as UNIX seconds, resulting in erroneous 1970 dates.
Recommendation systems like Draco~\cite{moritz2018formalizing} alleviate generic configuration challenges, but modeling the exponential spaces of all parameters, especially those for layered and composite charts (e.g., waterfall, pyramid), across different libraries remains impractical.

As LLM-based agents are increasingly used to generate specifications from natural language or demonstrations, overcoming this configuration barrier has become urgent. The challenge is twofold. \emph{From the user perspective}, even when agents produce visually effective charts, the resulting specifications are often verbose and \look{fragile, with hard-coded interdependent low-level parameters}. Small changes \look{such as} altering encodings or swapping chart types often break the code, forcing users to request full regeneration and incurring high token costs and latency that hinder interactive exploration. \emph{From the developer perspective}, building agents that consistently follow best practices across libraries remains difficult. Visualization grammars have long-tailed parameter spaces and use different vocabularies. Training data is also uneven across libraries. These factors make agents unreliable for less common chart types and libraries. Developers therefore face substantial challenges in training or instructing LLM agents to robustly generate correct and well-designed visualizations.

In this paper, we argue that the trade-off between fragile, high-quality charts and flexible, poorly designed defaults comes from a language limitation: existing visualization grammars lack an explicit way to specify the semantics of data. Many low-level parameters implicitly encode semantic decisions, such as parsing rules, meaningful baselines (e.g., zero), valid aggregations, and formatting. Without semantics as first-class objects, human authors and LLM agents must encode their intent indirectly through monolithic, library-specific parameters. Conversely, when semantics are omitted entirely, systems are forced to ``guess'' data meanings, often producing suboptimal or misleading designs.

To address these challenges, we introduce \ourrep, a semantics-driven intermediate language for visualization. Inspired by database research that leverages semantic types for data integration and cleaning~\cite{hulsebos2019sherlock,DBLP:journals/corr/abs-2105-00121}, our key insight is to formalize a hierarchical semantic type system as first-class objects in visualization specifications. Rather than requiring users or LLMs to manipulate low-level parameters, \ourrep explicitly represents data semantics as types to guide visualization design. \ourrep's \look{data-driven} compiler integrates data semantics, raw data representations, and chart specifications to automatically infer critical configurations, including scales, axes, aggregations, formatting, and layout. As a result, users only need to specify a chart type and field-to-channel mappings, while the system derives the remaining low-level details. This design aligns with users' preference for using simple chart-type-based approaches to describe visualization goals during data exploration~\cite{bako2022streamlining,bako2022understanding,mcnutt2021integrated}.

Designed as a library-agnostic intermediate language, \ourrep bridges high-level chart intent and executable rendering grammars. Like intermediate representations in general-purpose programming (e.g., LLVM~\cite{lattner2004llvm}), it does more than resolve semantics: the  compiler also optimizes chart specifications and generates output for the target library selected by the user, such as Vega-Lite, Apache ECharts, or Chart.js. For example, the same heatmap specification can be adapted to data density: a dense matrix yields compact cells, fewer ticks, and abbreviated labels, whereas a sparse matrix produces larger cells and more detailed annotations. By embedding such design knowledge into compiler passes, \ourrep reduces reliance on carefully designed prompting and ad hoc defaults, making specifications more robust and easier to revise during iterative exploration. Its flexible backend architecture also lets users switch rendering libraries without rewriting chart specifications. In our implementation, we realize backends for three libraries with substantially different specification models: Vega-Lite, Apache ECharts, and Chart.js.

Through a case analysis (\autoref{sec:evaluation}), we show that \ourrep can compile concise semantic specifications into detailed, backend-specific code. We also show that \ourrep is well suited to LLM-based generation: semantic types are often easy for AI agents to infer from data field names, value patterns, and common sense, enabling the model to produce high-quality charts from compact semantic inputs. Compared with directly generating library-level specifications, \ourrep also enables LLMs to produce much shorter programs while maintaining strong visual quality, as verified by LLM judges~\cite{chen2024viseval}. Our contributions are:
\begin{itemize}[leftmargin=*]\itemsep0pt
    \item We introduce \ourrep, an intermediate language for specifying charts using data semantics \look{and} simple visual encodings.
    \item We develop a compiler that resolves and optimizes low-level chart configurations holistically from semantic intent, transpiling them to different visualization backends.
    \item \look{We release \ourrep with compilers for Vega-Lite, Apache ECharts, and Chart.js, and evaluate its effectiveness through comprehensive chart galleries and LLM-generation experiments.}
\end{itemize}

 \section{Related Work}
\bpstart{Declarative Visualization Specification}
Modern chart authoring is fundamentally shaped by declarative visualization grammars such as Wilkinson’s \textit{Grammar of Graphics}~\cite{wilkinson2006grammar}, alongside implementations like ggplot2~\cite{valero2010ggplot2}, Vega~\cite{satyanarayan2015reactive}, and Vega-Lite~\cite{satyanarayan2016vega}. These grammars provide a structured space of encodings, composition operators, and interaction mechanisms. This foundation enables authors to systematically reason about design decisions and iteratively refine charts, vastly improving upon ad hoc imperative code. Simultaneously, widely adopted libraries like D3~\cite{bostock2011d3}, \look{Mascot.js~\cite{liu2024manipulable}}, Matplotlib~\cite{tosi2009matplotlib}, and Apache ECharts~\cite{li2018echarts} provide highly expressive rendering capabilities and extensive chart-type coverage. However, these systems (even high-level grammars like Vega-Lite) often demand low-level parameter configurations for marks, axes, scales, layouts, and styling to better reflect data semantics, because system defaults inferred from surface data representations can be fragile \look{and hard-coded design choices can be difficult to edit~\cite{snyder2023divi}}. \look{Our work addresses this gap; \ourrep leverages data semantics to manage low-level library-specific chart configuration details, and it leverages existing libraries for final rendering.}

\bpstart{Template-Based Visualization Authoring}
Code reuse through templating is a common practice in both general programming and visualization design~\cite{snyder2025challenges}. Many visualization tools (e.g., Excel) and grammars utilize templates or predefined ``chart types'' to accelerate the generation process, such as for waterfall, lollipop, and sunburst charts. However, traditional template-based approaches~\cite{mcnutt2021integrated,bako2022streamlining} are often very restrictive. Static templates tightly couple a chart's structural scaffold with its underlying data configuration, making them restrictive and difficult to adapt to changes in data type or cardinality. More configurable templates offer greater flexibility, but they often expose a large number of low-level parameters, which can be difficult for both users and LLM agents to use effectively. \ourrep can be viewed as a form of ``dynamic template'' that decouples visual structure from data semantics: semantic configurations are determined dynamically from the data, while structural composition is governed by reusable templates. In this way, \ourrep complements prior work on visualization reuse by enabling flexible templates for bespoke charts that can automatically adapt to diverse datasets.

\bpstart{LLM-based NL2VIS}
Recent advancements in LLM-based NL2Vis techniques~\cite{elshehaly2025designing,gao2015datatone} have enabled the translation of natural language (NL) queries into visualizations by generating grammar-specific code (e.g., Vega-Lite) or utilizing multi-stage pipelines. Systems such as Chat2VIS~\cite{maddigan2023chat2vis} and LIDA~\cite{dibia2023lida} demonstrate that general-purpose LLMs can autonomously select chart types and produce viable specifications. 
Subsequent research has focused on improving robustness: NL4DV-LLM~\cite{sah2024nl4dvllm} uses LLMs to infer analytic tasks and attribute roles, while VegaChat~\cite{hostnik2026vegachat} introduces automated heuristic corrections and schema validation to guide error recovery. Recent state-of-the-art systems use LLMs to directly generate low-level visualization code~\cite{wang2025data,brossier2026state}. Although these specifications can yield high-quality charts, they are often complex and fragile, with interdependent parameters---even small edits require full regeneration---which increases both latency and cost for exploration. LLM performance also remains uneven across libraries because training data is imbalanced~\cite{chen2024viseval}. In response, we introduce \ourrep, a semantic intermediate language that lets LLMs translate natural-language intent into a concise representation, while the compiler automatically handles low-level configuration and adapts the result to different visualization backends based on data semantics. 

\bpstart{Visualization Design Knowledge}
Research in visualization has established foundational principles of visual encoding. Early work by Bertin~\cite{bertin1983semiology} and Cleveland and McGill~\cite{cleveland1984graphical} demonstrated that the accuracy of information recovery depends on the alignment between data types and perceptual tasks.
Subsequent research has extended these insights to higher-level analytical tasks, such as interpreting aggregates, distributions, and correlations, while articulating specific perceptual guidelines. Examples include ``banking to 45$^\circ$'' for optimal aspect ratios~\cite{cleveland1993model,wang2017there, wang2018image}, selecting perceptually distinct hues for multi-class categorical data~\cite{gramazio2014relation, lu2020palettailor}, and addressing the sine illusion in streamgraph ordering~\cite{vanderplas2015signs, bu2020sinestream}. These studies underscore that effectiveness is not merely a matter of encoding choice, but a complex interaction between data semantics, layout, and human perception.

Embedding such design knowledge into practical systems remains difficult, especially in AI-based ones where LLM agents do not reliably follow best practices because of uneven training data, even with complex prompting. \ourrep addresses this issue by encoding design knowledge directly as deterministic optimization passes that systematically adjust visualization parameters. This separation lets LLMs focus on \emph{what} to visualize, while the compiler determines \emph{how} to visualize it well. Because \ourrep is library-agnostic, it also lowers the cost of transferring these design principles across practical rendering systems.


\bpstart{Recommendation systems} Automatically recommending visualizations from data or partial inputs can help users better explore data. Early systems like APT~\cite{mackinlay1986automating} and ShowMe~\cite{mackinlay2007show} relied on physical data types to select encodings, while modern systems like Draco~\cite{moritz2018formalizing} formalize design knowledge into strict constraints and preference rules to rank Vega-Lite charts. However, because low-level configuration spaces are large and difficult to cover exhaustively, existing recommendation engines are often restricted to a small set of common chart types with simplified parameterizations, limiting the exploration space. \ourrep complements these systems by abstracting away low-level configuration details and exposing semantic information directly. This allows recommendation engines to focus on higher-level design decisions, such as chart type, encoding composition, and analytical task fit, while relying on the compiler to handle downstream parameterization.

\bpstart{Semantic Types} Semantic type modeling has been widely studied in the database community for data cleaning and integration. Systems such as Sherlock~\cite{hulsebos2019sherlock} and Sato~\cite{zhang13sato} predict fine-grained concepts from tabular data using machine learning, while more recent approaches use LLMs~\cite{li2150llm} to further improve detection. Modern analytics platforms~\cite{GLooker,DBLP:journals/corr/abs-2105-00121} also increasingly deploy semantic layers to define entities and governed metrics, ensuring consistent aggregation and interpretation. These efforts highlight the value of explicit data semantics, but existing models are designed primarily for data preparation and are not fine-grained enough for visualization needs.
To address this gap, we introduce a visualization-oriented semantic type hierarchy. By treating semantic types as first-class objects, our compiler can constrain parameter choices and systematically generate higher-quality charts while reducing user configuration burden.

\section{Overview}

In this section, we first use an example to explain challenges for systematically handling low-level chart configurations in existing libraries. Then, we provide an overview of our language design and the compilation approach to address this problem.

\subsection{Motivating Examples}
\begin{figure}[h]
\centering
\begin{minipage}[c]{0.5\columnwidth}
\small
\setlength{\tabcolsep}{3pt}
\begin{tabular}{@{}cccc@{}}
\toprule
\textbf{Period} & \textbf{Week} & \textbf{AgeGroup} & \textbf{Users} \\
\midrule
202001 & wk1 & 18--24 & 15412 \\
202001 & wk1 & 25--34 & 24440 \\
202001 & wk1 & 35--44 & 20241 \\
\multicolumn{4}{c}{\dots} \\
202001 & wk2 & 18--24 & 16255 \\
\multicolumn{4}{c}{\dots} \\
202512 & wk4 & 60+    &  7449 \\
\bottomrule
\end{tabular}
\end{minipage}~
\begin{minipage}[c]{0.5\columnwidth}
  \includegraphics[width=\linewidth]{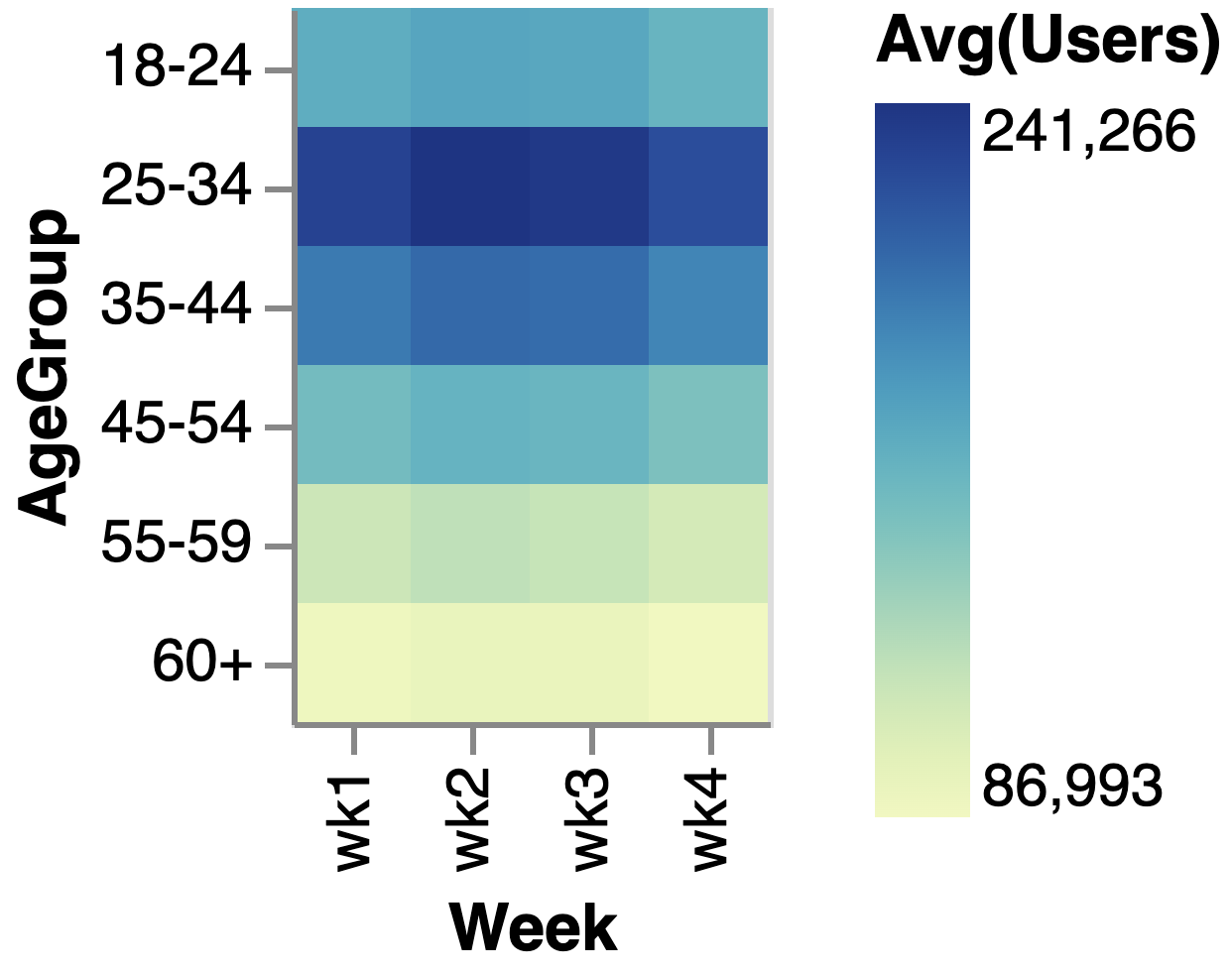}
\end{minipage}\vspace{-2mm}
\caption{Example user-engagement dataset and a simple heatmap.}\label{fig:scenario1}
\end{figure}

We use a user-engagement dataset from a consulting scenario as a running example. It contains four fields: \code{Period}, \code{Week}, \code{AgeGroup}, and \code{Users}. The dataset records active-user counts over 72 monthly periods (\code{202001} to \code{202512}), with each period further broken down by week (\code{wk1}--\code{wk4}) and age group (\code{18--24}, \code{25--34}, \dots, \code{60+}). Sample rows are shown in \Cref{fig:scenario1}.

\vspace{-0.8em}
\vspace{-0.8em}

From this dataset, the user can easily create a heatmap to compare week and user group trends: \code{Week} on the $x$-axis, \code{AgeGroup} on the $y$-axis, and \code{avg(Users)} as color. This chart is easy to create in Vega-Lite using the \code{rect} mark. 

\vspace{-0.8em}
\begin{minted}{json}
{ "mark": "rect", "encoding": { 
    "x": {"field": "Week"}, 
    "y": {"field": "AgeGroup"}, 
    "color": {"field": "Users", "aggregate": "mean"} } }
\end{minted}
\vspace{-0.8em}

\noindent Because the specification is short and the default primitive data types align perfectly with visual mappings, it is easy to adapt: with minor edits, the user can swap the $x$ and $y$ encodings or try a different chart type, such as a grouped bar chart. 

The analyst now wants a small conceptual edit: replace \code{Week} with \code{Period} on the $x$-axis and change the aggregation from \code{avg(Users)} to \code{sum(Users)}.
Ideally, the user should simply be able to swap the field and update the aggregate, an edit of only two tokens, and obtain a properly rendered chart. However, this is where the gap between primitive data types and \textbf{semantic types} becomes a critical hurdle. While \code{Period} values (e.g., \code{202001}) are stored as raw integers, their semantic type is a structured temporal unit (Year-Month). Because standard visualization grammars do not natively understand these rich semantic types, the user faces a dilemma: accept a simple specification that produces poor results, or write a verbose specification burdened with tightly coupled, low-level parameters.

\begin{figure}[h]
    \centering
    \includegraphics[width=0.8\linewidth]{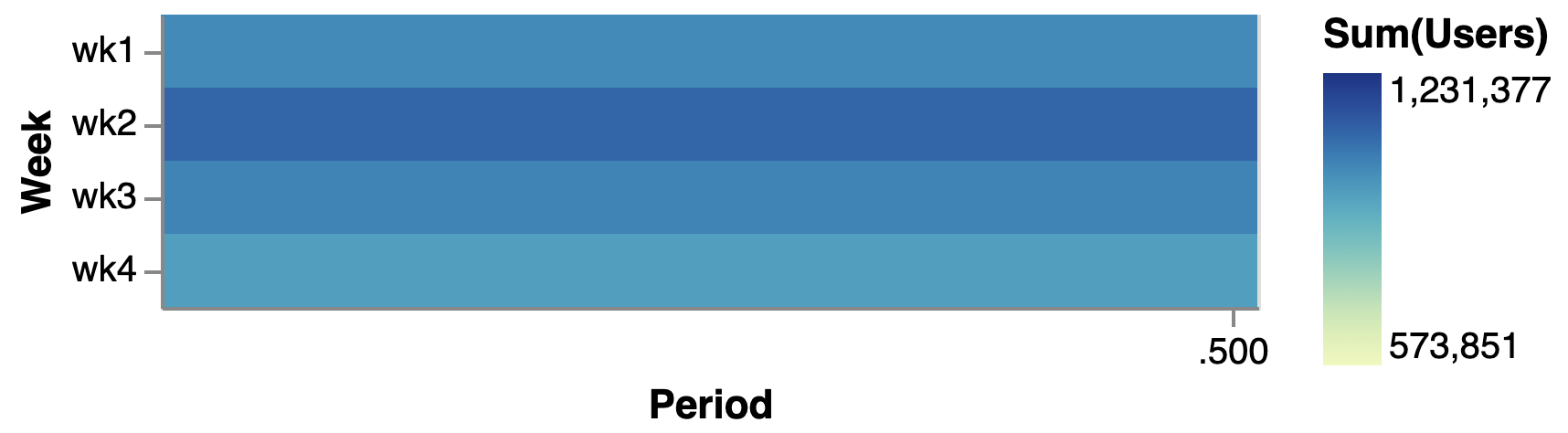}

    \includegraphics[width=\linewidth]{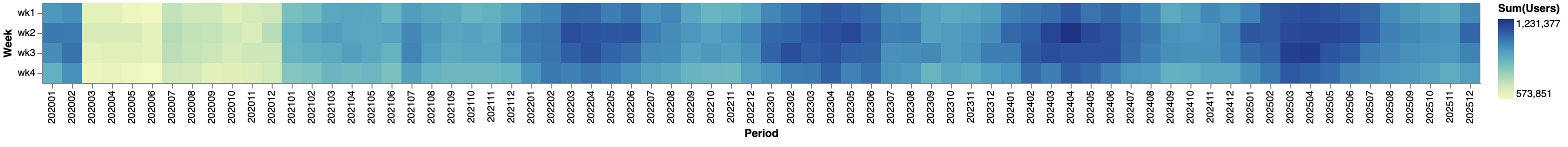}\vspace{-2mm}
    \caption{A simple specification fails in two different ways: temporal interpretation misparses the period field, while nominal interpretation produces an overcrowded layout.}
    \label{fig:scenario2-failures}
\end{figure}

\bpstart{Option 1: Simple but problematic specification}
A short specification no longer produces a satisfactory chart because the system misinterprets the underlying semantics of the data. If \code{Period} is naively cast as a temporal type directly from the integer \code{202001}, the system misparses the dates, resulting in an invalid timeline and wrong geometry (\Cref{fig:scenario2-failures}, top). Conversely, if it is treated as a nominal string, the chart blindly expands into 72 discrete categories (\Cref{fig:scenario2-failures}, bottom). Without understanding that \code{Period} here can simultaneously behave as discrete values to control tile configuration and a continuous temporal axis for parsing and axis formatting, the compiler cannot automatically decide a set of default parameters to produce a good visualization. 

\vspace{-0.8em}
\vspace{-1em}

\bpstart{Option 2: Detailed but fragile \look{specification}}
An expert can create a clean heatmap for \code{Period $\times$ Week $\times$ sum(Users)}. However, doing so requires coordinating a fragile set of low-level decisions that compensate for the system's lack of semantic awareness:
\begin{itemize}[leftmargin=*]\itemsep0pt
\item \emph{Manually enforcing temporal semantics:} Because the system sees \code{202001} merely as a raw integer, the expert must explicitly parse it using a \code{\%Y\%m} format and force the axis to be \code{temporal}. This explicit intervention prevents the chart from rendering arbitrary numbers and forces the use of time-based tick generation algorithms for readable labels.

\item \emph{Adapting physical geometry to accommodate high data cardinality:} 
\look{To reconcile the broken positional behavior, the expert must coordinate a fragile dual-treatment: treating \code{Period} as continuous for chronological parsing and axis tick generation, while managing it as discrete for heatmap tile alignment. This requires hard-coding interdependent layout parameters, specifically a 7px mark width and a 504px canvas span, and manually padding the axis domain bounds to \code{["2019-12-17T00:00:00", "2026-01-16T00:00:00"]}. This padding ensures that the first and last time-based ticks precisely align with the outer edges of the dense tiles without text overlap.}
\item \emph{Refining color domain:} Because \code{sum(Users)} encodes absolute quantitative magnitude, its visual encoding must remain mathematically proportional. The expert therefore fixes the color scale to a zero baseline, preventing auto-scaling from distorting the perceived ratios of user counts.
\end{itemize}
\begin{figure}[h]
    \centering
    \begin{minipage}{1\linewidth}
\begin{minted}[fontsize=\small]{json}
{ "transform": [{
    "calculate": "timeParse(toString(datum.Period), '%Y%m')",
    "as": "Date" }],
  "mark": {"type": "rect", "width": 7},
  "encoding": {
    "x": { "field": "Date", "type": "temporal",
      "scale": { 
        "nice": false,
        "domain": ["2019-12-17T00:00:00", "2026-01-16T00:00:00"] } },
    "y": { "field": "Week" },
    "color": {
      "field": "Users", "aggregate": "sum", 
      "scale": {"domainMin": 0} } },
  "config": { "view": {"continuousWidth": 504} }}
\end{minted}
    \end{minipage}
    \vspace{5pt}

    \begin{minipage}{1\linewidth}
        \centering
        \includegraphics[width=\linewidth]{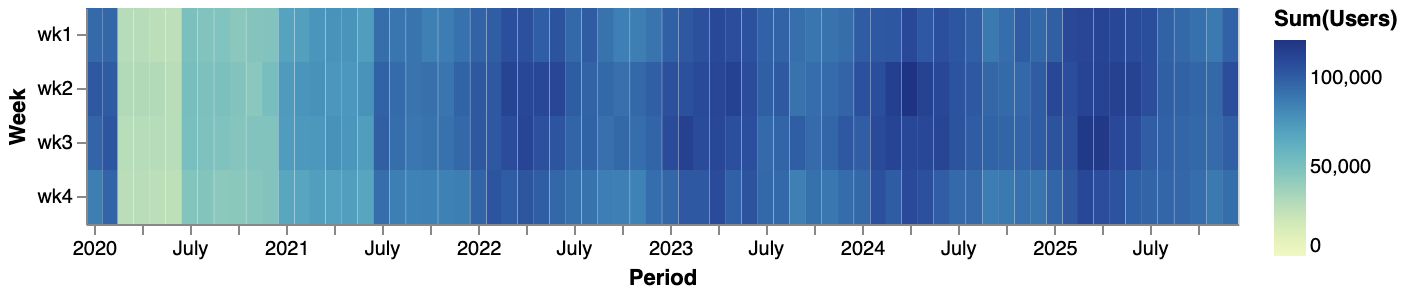}
    \end{minipage}
\vspace{-3mm}
    \caption{\look{An expert can produce a clean heatmap for \code{Period $\times$ Week $\times$ sum(Users)}, but only by coordinating tightly coupled low-level parameters, including parsing, scale, mark width, and layout.}}
    \label{fig:expert-heatmap-spec}\vspace{-6mm}
\end{figure}

\Cref{fig:expert-heatmap-spec} shows the resulting Vega-Lite specification and rendered heatmap. While effective, the solution is fragile because it depends on hard-coded, interdependent parameters that a semantics-aware system should ideally infer automatically. This limitation becomes apparent when the user makes a seemingly simple change, such as rotating the chart to \code{AgeGroup $\times$ Date $\times$ sum(Users)} and faceting by \code{Week}. The chart representation changes substantially: \code{Week} becomes a facet, \code{Date} moves to a vertical temporal axis, and \code{AgeGroup} becomes a horizontal categorical axis, so the previous configuration no longer transfers. The user must manually redefine layout, spacing, geometry, and cross-facet color consistency.

These challenges extend beyond this example and are especially pronounced in compositional visualizations such as pyramid or waterfall charts, where parameters must be coordinated within and across layers. The issue is further amplified across visualization libraries with different abstractions. For example, Vega-Lite controls layout through encodings, scales, and view properties, while ECharts organizes charts using series, grids, and component configurations. Even in the heatmap example, categorical spacing is handled through step size in Vega-Lite but through layout regions and paddings in ECharts. This diversity forces users and AI agents to reason about different parameterizations for the same visual intent. This scenario motivates the need for a unified, library-agnostic model that derives low-level specifications from high-level semantics.

\begin{figure*}[!t]
    \centering
    \includegraphics[width=\linewidth]{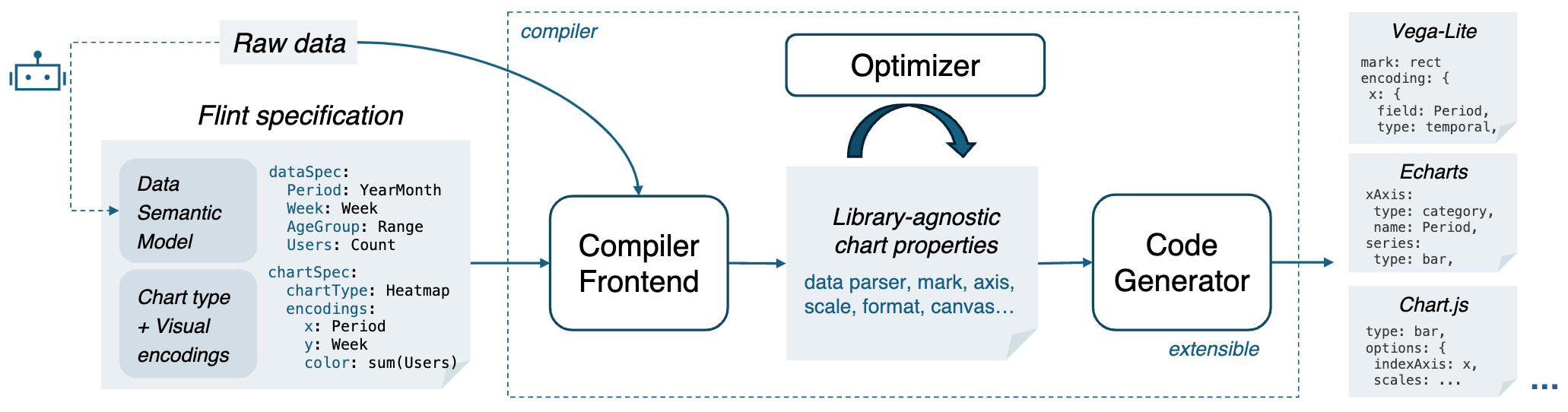}\vspace{-3mm}
    \caption{Overview of the \ourrep architecture. \ourrep employs a modular three-stage design: a \textit{compiler frontend} for translating user intents into library-agnostic properties, an \textit{optimization stage} that refines configurations, and \textit{extensible code generators} that produce library-native specs for desired rendering engines. LLM agents can infer data semantic specs from the raw data to reduce authoring efforts.}
    \label{fig:Architecture}\vspace{-6mm}
\end{figure*}

\subsection{System Architecture}
\label{sec:architecture-overview}

Our solution is to design an intermediate language that lets users specify visualization intent through high-level declarative representations while the compiler handles low-level configuration automatically across libraries. The specification includes the semantic structure of the data and the intended chart. The full specifications for the experts' heatmaps are shown here.

\vspace{-0.5em}
\begin{center}
\begin{minipage}[t]{0.3\linewidth}
\begin{minted}[fontsize=\small]{yaml}
dataSpec:
  Period: YearMonth
  Week: Week
  AgeGroup: Range
  Users: Count
\end{minted}
\end{minipage}
\begin{minipage}[t]{0.33\linewidth}
\begin{minted}[fontsize=\small]{yaml}
chartSpec:
  chartType: Heatmap
  encodings:
    x: Week
    y: AgeGroup
    color: avg(Users)
\end{minted}
\end{minipage}
\begin{minipage}[t]{0.33\linewidth}
\begin{minted}[fontsize=\small]{yaml}
chartSpec:
  chartType: Heatmap
  encodings:
    x: Period
    y: Week
    color: sum(Users)
\end{minted}
\end{minipage}
\end{center}
\vspace{-0.5em}
Such simple specs make chart adaptation easy: the user only updates the high-level chart intent, while the data specification remains unchanged. When the chart specification changes, the compiler combines semantic information with raw data characteristics to (re)infer the parameters required by the target library, freeing users from low-level specification management while still producing high-quality charts.

\Cref{fig:Architecture} shows our system architecture: a modular three-stage compilation pipeline inspired by the LLVM architecture~\cite{lattner2004llvm}. The key idea is to maintain and optimize a set of abstract visualization properties in a library-agnostic fashion that reflects data semantics before translating them into backend-specific executable specifications.

\bpstart{Compiler Frontend}
The pipeline begins with raw data and a user-defined \ourrep specification, including data semantic specs, the chart type, and visual encodings. The frontend resolves encoding properties from these inputs and represents them as a shared \emph{compilation context}. Properties produced from this stage capture core encoding decisions reflecting the semantic meaning of data, including data parsing, encoding type, formatting behavior, aggregation, scale, domain, ordering, and midpoints.

\bpstart{Optimizer}
The optimizer enriches this compilation context with layout decisions, coordinating the sizes of visual elements with the available canvas while accounting for data cardinality, encoding types, grouping, and faceting. It first resolves local layout properties for each axis, such as axis span, band step, padding, group width, radius, and facet panel size, and then refines them globally to resolve dependencies like aspect ratio and facet wraps. Thus, this stage augments the compilation context with \emph{optimized layout properties} that ensure the chart will be readable and well-proportioned, avoiding overcrowded, distorted, or excessively sparse layouts.

\bpstart{Code Generator} Finally, the code generator decides final chart-specific properties and translates the optimized compilation context into an executable specification for a target visualization library (e.g., {Vega-Lite}, {ECharts}, or {Chart.js}). The code generator handles the diverse vocabulary and programming models used by different libraries. The backend is modular and extensible: each library provides its own templates and compilation parameters, allowing new libraries to be added without changing the frontend or optimization stages. The output is a library-native specification ready for rendering.



\section{The \ourrep language}
\label{sec:language-design}

In this section, we present the design of our intermediate language \ourrep. \ourrep represents visualization-relevant data semantics as first-class objects in the specification, so that low-level chart decisions can be inferred systematically from them. Our design follows these two principles:
\begin{enumerate}[leftmargin=*]
\item We decouple \emph{what} data means from \emph{how} it is mapped to visual properties. Instead of relying solely on coarse encoding-oriented data types such as quantitative, nominal, temporal, and ordinal, we introduce fine-grained, encoding-agnostic semantic types that model the meaning of data fields directly. This builds a reusable semantic representation that can support multiple visualizations of the same data.
\item We adopt a chart-type-based approach to visualization specification, where the compiler manages layering, composition, and instance-specific parameters using configuration decisions inferred from data semantics. This keeps specifications concise while supporting expressive visualizations and simplifying adaptation across chart instances.
\end{enumerate}

\subsection{Hierarchical data semantic model}
Our goal is to formalize a semantic type system that lets users formally express data meaning that can guide visualization design. Although semantic types have been used in database research for data mining tasks~\cite{hulsebos2019sherlock,DBLP:journals/corr/abs-2105-00121}, existing definitions are often too coarse for visualization. For example, two columns associated with month may require very different visual treatments if one stores month-of-year codes (\code{1--12}) and the other stores year--month timestamps (e.g., \code{2001-01} to \code{2010-12}). The former is cyclically ordered discrete representation, whereas the latter forms a chronological sequence suits continuous temporal scales. Treating both as the same \code{Temporal} or \code{Month} type obscures visualization-relevant distinctions. At the same time, overly fine-grained labels such as \code{MonthName}, \code{MonthNumber}, or \code{UnixTimestamp} increase specification complexity without improving visualization intent, making schemas brittle.

\begin{table}[t]
  \centering
  \caption{Data semantics hierarchy examples.} \vspace{-3mm}
  \label{tab:type-registry}
  \small
  \setlength{\tabcolsep}{6pt}
  \renewcommand{\arraystretch}{1.1}
  \begin{tabular}{@{}lll@{}}
    \toprule
    \textbf{L1: Domain} & \textbf{L2: Family} & \textbf{L3: Types} \\
    \midrule
    \multirow{3}{*}{Temporal}
      & DateTime    & Date | Time | Timestamp | ...\\
      & DateGranule & Year | Quarter | Month | YearMonth | ... \\
      & Duration    & Duration \\
    \midrule
    \multirow{3}{*}{Categorical}
      & Entity & Category | Name \\
      & Coded & Status | Boolean | Direction \\
      & Binned & Range \\
    \midrule
    \multirow{5}{*}{Measure}
      & Amount         & Revenue | Cost | Price | ... \\
      & Proportion     & Percentage \\
      & Rating & Score | Rating \\
      & Signed  & Profit | Correlation | Delta | ... \\
      & Generic & Quantity | Number | Temperature | ...\\
    \midrule
    \multirow{1}{*}{Discrete}
      & Rank & Rank \\
    \midrule
        Identifier
      & ID          & ID \\
    \midrule
    \multirow{2}{*}{Geographic}
      & Coordinate & Latitude | Longitude \\
      & GeoPlace & Country | State | City |   ZipCode | ...\\
    \bottomrule
  \end{tabular}\vspace{-4mm}
\end{table}

Our key insight here is to formalize a semantic type system based on \emph{how data meaning influences visualization behavior}, and to organize these types hierarchically so that each level captures a different class of compilation decisions. We formalize semantic types as a three-level hierarchy: Level~1 determines how raw values should be interpreted and parsed; Level~2 captures scale- and transformation-level semantics, such as aggregation behavior, zero interpretation, and domain structure; and Level~3 refines type-specific presentation choices, such as formatting, ordering, and other display defaults. \look{This hierarchy follows an inheritance structure: lower-level types inherit the compilation decisions made by their ancestors and then add more specific constraints when needed. For example, a Level~3 type such as \texttt{YearMonth} inherits the temporal parsing behavior from its Level~1 \texttt{Temporal} domain and temporal scale behavior from its Level~2 family, while refining presentation-level details such as date formatting and tick labeling. Thus, deeper levels do not alter the higher-level interpretation of the data; they specialize it for more reliable visualization generation.}

\begin{itemize}[leftmargin=*]\itemsep0pt
\item \emph{Level 1: Semantic Domain.}  
We first group semantic types based on \emph{how raw data should be parsed}. 
\look{Our design leverages this level to determine the core parser class (e.g., temporal or numeric) to ensure correctness of data processing.}
For example, a column with values \code{202001}, \code{202002}, \code{202003} might represent year--month timestamps, or quantities such as revenue or account identifiers depending on their column names and contexts. If a system relies only on automatic parser tests, it may silently accept revenue as a datetime object and configure it into a temporal axis; they may also turn a numerical field representing sales with missing values into nominal just because the column has values represented as ``\code{N/A}''. With Level 1 semantic type guidance, the field can be parsed and interpreted with greater confidence.

\item \emph{Level 2: Semantic Family.}
We next group semantic types based on \emph{how data behaves with respect to key encoding properties} like aggregation, zero, and domain shape. For example, despite both revenue and percentage are numeric measures that should be processed by numerical parsers (per L1 decision), they imply different aggregation and domain properties. Revenue is an \textit{Amount}: it is additive, zero denotes the absence of the measured quantity, and its domain is typically open-ended. Percentage, in contrast, is a \textit{Proportion}: it is intensive and often has either a bounded or context-dependent domain. These distinctions affect chart configurations: when they are used as $y$-axis of a bar chart but color encoding is not provided, an \textit{Amount} field such as revenue should typically default to summation if there are duplicate values, whereas a \textit{Proportion} field such as percentage should default to average. Other example properties determined in L2 are whether including zero in the domain should be enforced, optional or disabled; whether the scale domain should be fixed or data-dependent.

\item \emph{Level 3: Semantic Type.} 
\look{Finally, L3 captures fine-grained semantic types that define presentation decisions, }
such as diverging, formatting, ordering, and ticks.
\look{For example, a \code{Rank} field should be displayed on a reversed axis so that rank \code{1} at the top.} Similarly, in a dot plot that compares the longitudinal positions of cities along an east-west axis, a \code{Longitude} field should respect its closed semantic domain $[-180,180]$ rather than being displayed on an arbitrarily padded axis like $[-200,200]$. Such fine-grained distinctions provide more consistent visual presentation.
\end{itemize}

\noindent \look{In \ourrep, this hierarchy is implemented as a global semantic type registry containing 6 L1 domains, 15 L2 families, and 44 L3 types (\autoref{tab:type-registry}). The registry operationalizes the inheritance structure described above: each L3 leaf type inherits parser decisions from its L1 domain and encoding properties from its L2 family, then contributes type-specific presentation refinements such as domain shape, divergence, formatting, and ordering. This structure also supports progressive disclosure. LLM agents can use the full L3 registry for precise visualization configuration, while human authors can specify only intuitive L1/L2 types and let the compiler fill in best-effort defaults. \autoref{tab:semantic-properties-example} illustrates how concrete leaf types populate this compilation registry.}

\begin{table*}[ht]
\centering
\caption{Illustrative semantic type assignments across the seven abstract properties.}\vspace{-2mm}
\label{tab:semantic-properties-example}
\scriptsize
\begin{tabular}{@{}l|lllllll@{}}
\toprule
\textbf{Semantic Type}
& \textbf{Parser}
& \textbf{Encoding Candidates}
& \textbf{Zero Class}
& \textbf{Agg. Role}
& \textbf{Domain Shape}
& \textbf{Diverging}
& \textbf{Formatter Ex.} \\
\midrule
YearMonth  & temporal & temporal / ordinal        & none       & dimension        & open    & none        & \texttt{\%Y-\%m} \\
Month      & temporal & ordinal                   & none       & dimension        & cyclic  & none        & Jan / 1,2,3\\
Percentage & numeric  & quantitative              & contextual & intensive        & bounded & none        & \texttt{12.3\%} \\
Revenue    & numeric  & quantitative              & meaningful & additive         & open    & none        & \texttt{\$12.4K} \\
Profit     & numeric  & quantitative              & meaningful & signed-additive  & open    & conditional & \texttt{+\$2.1K} \\
Latitude   & numeric  & geographic                & arbitrary  & dimension        & fixed   & none        & \texttt{37.77} \\
Status     & string   & nominal                   & none       & dimension        & open    & none        & Active \\
Direction  & string   & ordinal / nominal         & none       & dimension        & cyclic  & none        & NE \\
\bottomrule
\end{tabular}\vspace{-6mm}
\end{table*}

Besides conceptual clarity, the hierarchical design provides direct engineering benefits, especially \emph{reusability} and \emph{extensibility}. For example, instead of building one set of parser for each semantic type, we only need to maintain a core set of parsers for each semantic domain. If a developer wants to introduce a new semantic type \code{FiscalQuarter} to support specialized labeling and formatting for financial use cases, they only need to implement the L3 semantics while reusing others based on parent types. In this way, the hierarchy reduces implementation redundancy, keeps the taxonomy maintainable, and allows new semantic types to be added without ad hoc parsing and visualization rules.

Note that semantic types do not replace the role of raw data representations \look{(i.e., the actual input data values)} in visualization. Rather, they are metadata of raw values that gives the compiler actionable guidance on how to appropriately interpret data for encodings, chart, and formatting defaults. In the next sections, we show how semantic types are complemented by data-specific semantic attributes, such as \code{intrinsic\_domain} and \code{mid}, and how these specifications are transformed into detailed, semantically aligned chart configurations.

\subsection{The language}

Building on the data semantics model, \ourrep defines a visualization specification in two separate sections: (1) a \code{dataSpec} that describes the semantic representation of data fields and (2) a \code{chartSpec} section that describes the intended visual mapping. \Cref{fig:grammar} illustrates the grammar of this visualization language. 
Because the data specification captures the semantics of the underlying fields independently of charts, it only needs to be created once and can be reused for charts of the same dataset. In practice, users do not need to author all of this information manually, as much of the data specification can be inferred by an LLM from column names, observed values, and surrounding context~\cite{hulsebos2019sherlock}. \look{Although data semantics and chart specifications are conceptually separated, they can still be authored together in a single \ourrep specification to reduce cognitive load during manual authoring. In LLM-powered interactive systems, the \code{dataSpec} can be inferred or cached when data is uploaded, making it easy for users to adapt the \code{chartSpec} when creating subsequent visualizations.}

\begin{figure}[t]
\centering
\begin{minted}[fontsize=\small]{yaml}
dataSpec:
  <field>:
    semanticType: ...
    intrinsicDomain: null | [<min>, <max>] | [<v1>, <v2>, ...]
    mid: null | <value>
chartSpec:
  chartType: Scatter | Bar | Grouped Bar | Line | Area | Heatmap | Radar | Pyramid | ...
  encodings:
    <channel>:
      field: <field>
      aggregate: null | sum | average | min | max | unique
      overrides: ...
\end{minted}
\vspace{-12pt}\vspace{-2mm}
\caption{Schema of \ourrep. {dataSpec} defines field semantics; {chartSpec} defines the chart type and simple encodings.}\vspace{-6mm}
\label{fig:grammar}
\label{fig:spec-schema}
\end{figure}

\bpstart{Data Specification} In \code{dataSpec} (\autoref{fig:grammar}), in addition to a semantic type, users can provide optional properties such as an \code{intrinsic domain} and \code{midpoint} to complement a field's semantic description. Semantic type captures the general meaning of a field, but it does not fully determine its intended representation for visualization. For example, a field with semantic type \textit{Rating} may use a 1--5 scale, a 0--100 scale, or an ordered discrete domain such as \code{[strong reject, weak reject, weak accept, strong accept]}, even if the observed table values cover only part of that range. Likewise, \code{accuracy} and \code{recall} are naturally bounded to $[0,100]$ (or $[0,1]$), whereas other percentage-valued fields may not imply the same closed range. Ordered fields such as age groups require an intrinsic domain to preserve the intended order rather than a lexicographic order \code{[<18, >60, 18--29, 30--44, 45--60]}. The \code{midpoint} property specifies a meaningful reference value, such as $0$ for signed scores or a neutral baseline for sentiment, which may differ from the numerical center of the domain (e.g., $32^\circ$F for Fahrenheit temperatures). These properties are helpful for visual presentations: intrinsic domains support correct axis ranges and ordering, while midpoints enable centered color scales and diverging encodings. Together, \look{these optional properties} complement semantic types to constrain chart configurations.

\bpstart{Chart Specification}
The \code{chartSpec} component describes the intended visualization. The user first chooses a \code{chartType}, which selects a visual template such as \code{Scatter}, \code{Line}, \code{Bar}, or \code{Heatmap}, and then specifies an \code{encodings} map that binds visual channels to fields or aggregated fields, such as $\code{x} \mapsto \code{Month}$, $\code{y} \mapsto \code{sum(Revenue)}$, or $\code{color} \mapsto \code{Region}$.

We adopt a chart type-based specification so that users can express a wide range of visualizations, including radar charts, pyramids, waterfall charts, and candlestick charts, without explicitly constructing them through composition algebra~\cite{satyanarayan2016vega}. This design matches \ourrep's role as an intermediate language centered on semantic intent rather than bespoke visual construction. At the same time, \ourrep remains expressive through extensible chart templates, allowing it to support new visualization forms without expanding the core language. Unlike traditional static templates, which are often rigid with respect to data types, cardinalities, or formats, these compiler-backed templates adapt to the data and semantics in the specification, automatically resolving chart-specific parameter dependencies to produce high-quality visualizations.

We deliberately keep the chart language simple by omitting low-level view configuration such as axes, scales, and formatting, leaving these decisions to the compiler (\autoref{sec:compiler}). At the same time, \ourrep preserves flexibility through optional chart properties and encoding-level overrides, which allow users to steer common chart behaviors without editing low-level generated code directly. For example, users may specify a map center for geographic charts or override the sort direction of the $x$-axis in a bar chart from ascending to descending. Finally, because the \ourrep compiler produces library-native code, users who want additional customization can further edit the compiled specification.

\look{
\medskip
\noindent\emph{Remarks.} Flint currently supports static chart specifications and does not expose interactive visualization features, such as brushing, zooming or cross-filtering, as first-class constructs. When these interactions are supported by backend libraries, users who need them can add or modify interactive components in the downstream backend specification generated by Flint.
In addition, Flint’s current semantic type design assumes a one-to-one mapping between data fields and semantic types. This can be insufficient for fields that admit multiple possible interpretations. We consider first-class support for interaction and ambiguous semantic types as future extensions to Flint’s core language. 

}

\section{\ourrep-Compiler}\label{sec:compiler}

\begin{figure}[t]
    \centering
    \includegraphics[width=\linewidth]{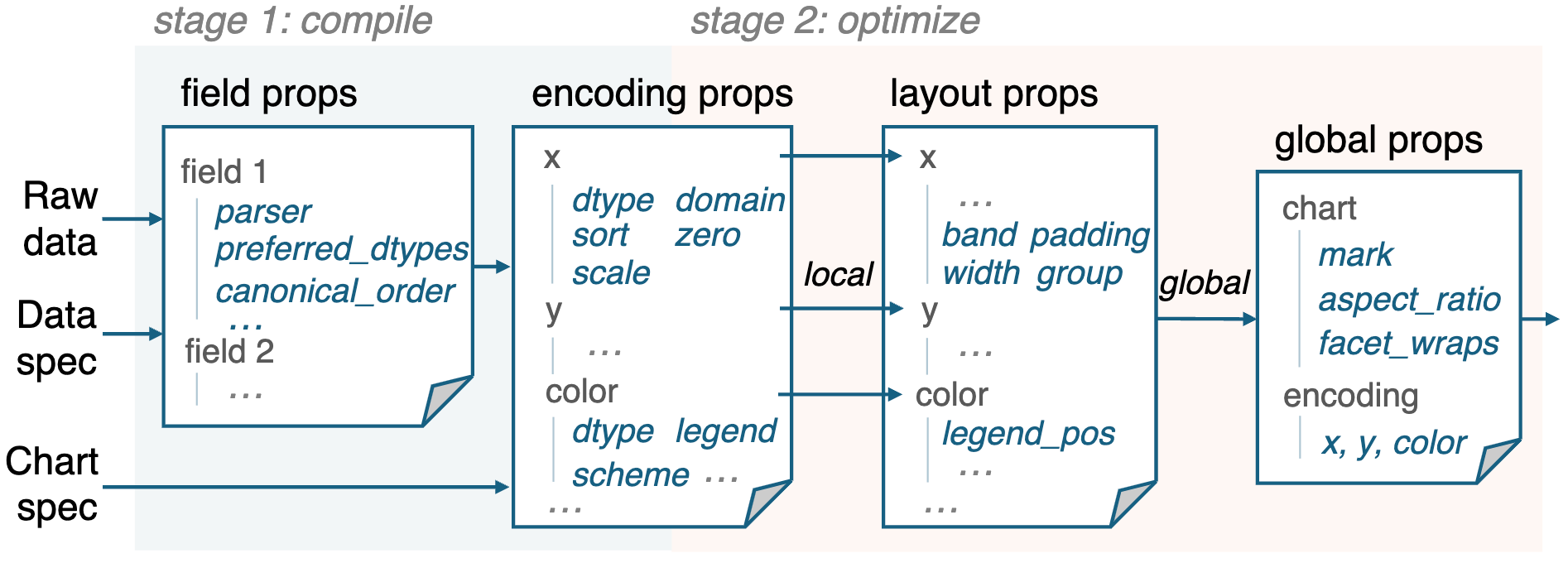}\vspace{-2mm}
    \caption{The compilation and optimization workflow. The system compiles raw inputs into intermediate {field} and {encoding properties}, then optimizes local and global properties to refine chart sizing.}
    \label{fig:compilation}\vspace{-6mm}
\end{figure}

As described in \autoref{sec:architecture-overview}, the compiler maintains a shared compilation context derived from the data and chart specification inputs, and progressively refines this context across three stages: semantics resolution, optimization, and code generation. Each stage resolves a different subset of visualization properties needed for high-quality chart design, namely \emph{encoding properties}, \emph{layout properties}, and \emph{template-specific properties}. \look{Flint treats data as part of the program: the compiler uses data characteristics, such as cardinality and distribution, alongside the specification during optimization. When the input data changes, Flint triggers a synchronous recompilation pass to adaptively recompute the visual layout.} \autoref{fig:compilation} provides an overview of this pipeline's first two stages.

\subsection{Stage 1: Semantics Resolution}
In this stage, the compiler frontend resolves the chart \emph{encoding properties} that capture the meaning of the data, including parsing, encoding type, formatting, scale type, tick constraints, axis reversal, stacking behavior, color semantics, and sort order. It does so by first resolving data ambiguity using both semantic types and raw data representations, and then grounds the decisions into chart-specific encoding properties.


The first step derives a set of \textit{field properties} that characterize each field independently of chart context. This step resolves the many-to-many correspondence between semantic types and raw representations described in \autoref{sec:language-design}. Guided by the semantic type, the compiler first selects an appropriate parser and then resolves the detailed field properties. Semantic guidance is crucial here, as it allows the compiler to disambiguate these representations and determine reusable semantics, such as whether the field can support temporal or ordinal treatment in later stages. This step also helps the compiler to resolve mismatches between semantic types and physical data information (\autoref{app:semantic-error-resolution}).

Next, the compiler derives \textit{encoding properties} for particular visual channels from field properties by adding chart contexts. For example, in the heatmap of \autoref{fig:expert-heatmap-spec}, placing \code{YearMonth} on the $x$-axis yields a temporal positional encoding with time formatting and an ordered temporal scale, but with \code{domainPadding=0} to accommodate discrete tiles. The same field may lead to different channel-specific decisions in other contexts; for instance, when used on the color channel of a line chart, \code{YearMonth} may be treated as an ordered or categorical progression rather than as a temporal positional axis depending on cardinality. Guided by the semantic type throughout, this step ensures that channel encodings remain semantically aligned, avoiding errors such as treating \code{YearMonth} as quantitative values or applying inappropriate aggregation to fields such as accuracy.

\subsection{Stage 2: Optimization} 

In the second stage, the compiler optimizes chart layout for the available display space. Users provide a preferred chart size and a maximum allowable area (e.g., preferred: $300 \times 300$, max: $800 \times 600$), and the compiler resolves \emph{layout properties} so the chart fits the canvas gracefully. It does this by balancing compression of individual visual components with expansion of the canvas within the allowed range. The properties considered include axis span, discrete step size, padding, group width, inter-group spacing, facet panel size, facet row and column counts, and, for radial layouts, radius and angular allocation. Without this stage, default settings can easily produce overflow, excessive density, or distorted layouts (e.g., the heatmap in \autoref{fig:scenario2-failures}). We formulate this stage as a physics-based optimization process: it first solves local parameters for each dimension, then coordinates these decisions to handle global properties such as aspect ratio.

\bpstart{Local optimizations} 
The optimizer first determines the size of each layout dimension that affects chart sizing, including the $x$- and $y$-axes, \code{group} (e.g., grouped bars), \code{radius} (for pie and radial charts), and \code{row}/\code{column} in faceted layouts. Each dimension acts as a 1D flexible container capable of expanding within a bounded budget. To ensure readability without relying on rigid static breakpoints, we adopt constraint-based approaches~\cite{hoffswell2020techniques, schottler2024constraint} to dynamically control element sizes. Depending on the encoding type, the optimizer resolves spatial demand as follows:
 \begin{itemize}[leftmargin=*]\itemsep0pt
 \item \emph{Discrete layouts.} For items occupying discrete slots (e.g., bars or heatmap cells), each mark acts as a compressible unit constrained by a preferred and minimum readable size. If total spatial demand exceeds the available span, the optimizer resolves the conflict by expanding the layout or compressing items until all minimum size constraints are met. 
 \item \emph{Continuous layouts.} For continuous axes (e.g., scatterplots or line charts), constraints are defined by visual density and allowable mark overlap. When high spatial density causes marks to violate these overlap constraints, the optimizer resolves the conflict by stretching the canvas dimensions to reduce crowding and ensure visual separation. 
 \end{itemize}

\noindent For nested charts (e.g., grouped bars, strip plots, small multiples), constraints are applied hierarchically. Spatial demand is first resolved for items within each group, followed by a joint optimization of how the groups themselves pack into the broader canvas budget. This ensures individual marks remain readable without violating the overarching limits of the stretched canvas.

\bpstart{Global optimizations} Building on local decisions, the optimizer resolves overarching chart-level properties, primarily aspect ratio (AR). For continuous charts, it blends an AR derived from the ``banking to 45$^\circ$'' principle~\cite{cleveland1993banking,heer2006multiscale,wang2017there,wang2018image} with locally stretched width and height, balancing visual interpretability against density reduction. For banded charts, it shrinks the orthogonal dimension to preserve proportional mark sizes and prevent excessive elongation. Second, it optimizes facet wrapping by balancing overall compactness against subplot legibility. Finally, for non-axis-bound charts (e.g., treemaps, gauge charts, funnel charts), it calculates canvas size directly from visual component counts. These final decisions yield an optimized compilation context ready for code generation. \look{For advanced users who wish to customize the compiler's layout behavior, Flint accepts compilation options such as stretch ratio, canvas size, base chart size, and minimum step size. These parameters are passed to the compiler at runtime and combined with data characteristics and chart design to constrain layout optimization, while preserving the same high-level visualization specification.}

\subsection{Stage 3: Code generation}
Finally, backend code generators translate the optimized, library-agnostic chart properties into library-native specs. They first resolve \emph{chart-specific properties} that depend on the selected chart template. Each backend provides a collection of \emph{dynamic chart templates}, where each template provides a chart skeleton with an instantiation function that consumes the compilation context, finalizes instance-specific properties, and emits a specification in the target library's native vocabulary. For example, a bubble chart template may adjust the pixel size range of bubbles based on data density, while a radar chart template may normalize radial domains using intrinsic field domains. The same optimized layout can therefore be realized differently across libraries: a Vega-Lite template may emit \texttt{width: \{step: ...\}} together with scale padding, whereas an ECharts template may use series-level properties such as \texttt{barCategoryGap}, \texttt{barGap}, and \texttt{barWidth}. This dynamic-template design allows each chart instance to inspect the full compilation context and adapt accordingly, improving flexibility for compositional charts. Overall, the compiler maintains a clear separation between optimization and code generation: chart intent is determined once, and each backend implements it according to the conventions of its target library.
\section{Evaluation}
\label{sec:evaluation}
In this section, we first demonstrate \ourrep's expressiveness and flexibility for human authoring through concrete examples, followed by an evaluation of its benefits for LLM agents in NL2VIS tasks. To facilitate this, we implemented \ourrep as a JavaScript library that currently supports three backends, Vega-Lite, ECharts, and Chart.js, and provides 30 chart types, including variants of scatter, bar, line, pie, density, maps and compositional charts. A gallery is shown in \autoref{app:gallery}.  We also integrate the library into the visualization pipeline of the open-source data analysis tool Data Formulator,\footnote{\url{https://data-formulator.ai}} where AI agents generate flexible \ourrep specs (rather than Vega-Lite directly) to facilitate exploration.


\subsection{Case Study} 
\label{sec:casestudy}

To demonstrate the effectiveness of \ourrep in supporting rapid, iterative data exploration, we conducted a case study using a synthetic dataset representing the 2025 global gaming market. The dataset tracks 24 games across 6 platform types (\code{gameType}) and 4 regions (\code{region}). The exploration process follows three phases, showing how \ourrep resolves the trade-off between specification simplicity and design quality. The example is shown in \autoref{fig:case-study}.

\begin{itemize}[leftmargin=*, labelsep=0pt, label={}]
\item \emph{Phase 1: Quick overview.}
The analyst begins by examining user trends across game types and regions using Vega-Lite. In traditional systems, this usually requires manual data and chart configuration to avoid misleading defaults: the \code{period} field could be parsed and formatted as integers, \code{regions} could be sorted alphabetically (\code{E,N,S,W}), and default subplot widths could become too large for four regions. With \ourrep, the compiler uses \code{dataSpec}, inferred by an LLM agent from the raw data, to configure the chart. With a simple \emph{chartSpec}, the analyst obtains a well-formatted multi-faceted visualization with optimized $x$- and $y$-axis labels, adapted chart widths, and semantically ordered facets, as shown in \autoref{fig:case-study}-d.

\item \emph{Phase 2: Exploring different visualizations.}
The analyst now wants to explore different data field combinations to better understand regional and user growth trends. Traditionally, as the visualization changes, the user must reconfigure the chart to accommodate new encoding properties and data cardinality: (1) when creating a grouped bar chart to compare user numbers across game types by month, the analyst must remap \code{period} from a temporal to a discrete axis and reconcile bar widths and axis lengths to avoid skewed chart proportions (\autoref{fig:case-study}-e); (2) when creating a waterfall chart to show user flux (\autoref{fig:case-study}-f), they must explicitly configure bar starting points to align adjacent bars; (3) when using \code{newUsers} in a heatmap (\autoref{fig:case-study}-g) to show each game's popularity trend over time, they must apply a diverging color scale with midpoint $0$ to reflect ``delta'' rather than relying on default linear color (e.g., blues) and revise $x$-labels to accommodate longer names. These reconfiguration steps accumulate and slow down analysis.
\ourrep eliminates this burden: the analyst simply changes the chart type and swaps encodings in the \code{chartSpec}, and the compiler reuses the \code{dataSpec} from the first step to automatically infer all downstream parameters to produce high-quality charts.

\item \emph{Phase 3: Switching backends.}
In the final phase, the analyst explores the user distribution hierarchy (\code{Region} / \code{GameType} / \code{Game}) using a sunburst chart. Since Vega-Lite does not support interactive sunburst charts, the analyst needs to use ECharts instead. However, the two libraries have very different programming models, so manually porting a chart can be costly. With \ourrep, the analyst only needs to provide a simple sunburst chart spec (\autoref{fig:case-study}-h) and switch the backend, thanks to \ourrep's library-agnostic design. If the analyst later wants to render any of the earlier Vega-Lite charts in ECharts to take advantage of ECharts' animation features, they can simply switch the backend without modifying the specification itself.
\end{itemize}

\noindent This exploration highlights three key advantages of \ourrep: (1) \emph{concise}---specifications were on average 85\% shorter than native backend code; (2) \emph{robust}---explicit semantic types prevent common design errors such as incorrectly formatted dates, invalid stacking of non-additive measures or diverging colors at the domain midpoint instead of zero; and (3) \emph{portable}---the same intent was seamlessly compiled into Vega-Lite for initial sketches and ECharts for final presentation.

\subsection{LLM Chart Generation}
To evaluate whether \ourrep is LLM-friendly, we compare the \emph{Flint} agent (Vega-Lite backend) with a baseline \emph{DirectVL} agent that outputs Vega-Lite directly from datasets and natural language queries. We follow the VisEval~\cite{chen2024viseval} protocol, which uses a vision-language model to assess charts generated from natural language queries. We exclude the standard VisEval dataset due to its SQL-focused complexity and limited visual diversity: 84\% of 1,150 cases are bar or pie charts (7 types total), 91\% use only two visual encodings, and tables are very small (median 4 rows, 86\% $\le$10). Instead, we evaluate on the more complex \textit{TidyTuesday} 2025 datasets.\footnote{\url{https://github.com/rfordatascience/tidytuesday}}
To address the lack of annotated queries, we introduce an automated three-stage LLM pipeline: a \emph{Question Generator} synthesizes diverse analytical queries~\cite{DBLP:journals/corr/abs-2509-21825}, a \emph{Chart} agent generates the visualizations using either the induced Flint-agent or the DirectVL-agent, and a vision-language \emph{Grader} evaluates the rendered images using the VisEval evaluator. Following VisEval, outputs are scored 0–5 on Relevance, Chart Errors, Clarity, and Design Quality, for a maximum of 20 points. Additional details on prompts and agent design are provided in \autoref{app:llm-eval-setup}.

\begin{table}[t]
\centering
\caption{Comparison of overall scores between \ourrep and {DirectVL} agents. Values show the number and percentage of wins/ties.} \vspace{-2mm}
\label{tab:eval-pairwise}
\scalebox{0.96}{
\begin{tabular}{lccccc}
\toprule
Model  & \#Flint wins & \#Tie & \#VL wins & $p$ \\
\midrule
GPT-5.1  & 129 (41\%) & 65 (21\%) & 118 (38\%) & 0.05 \\
GPT-5-mini  & 140 (45\%) & 65 (21\%) & 106 (34\%) & 0.001 \\
GPT-4.1  & 135 (43\%) & 70 (23\%) & 106 (34\%) & 0.004 \\
\bottomrule
\end{tabular}}\vspace{-6mm}
\end{table}

\bpstart{Observations} 
The experiment yields 315 questions (5 per dataset) across 10 chart types: Bar (18.7\%), Lollipop (15.6\%), Grouped Bar (14.6\%), Line (14.3\%), Boxplot (13.7\%), Scatter (11.4\%), Heatmap (7.0\%), Pie (2.5\%), Map (1.3\%), and Area (1.0\%). The median number of output rows is 37 across agents. \Cref{tab:eval-pairwise} shows pairwise comparisons: \ourrep wins more often than DirectVL, with statistical significance on GPT-5-mini ($p{=}0.001$) and GPT-4.1 ($p{=}0.004$), and borderline significance on GPT-5.1 ($p{=}0.05$). The advantage grows as model capability decreases, showing that embedding structured visualization knowledge into the language yields more reliable chart generation than unconstrained Vega-Lite, while remaining simple and flexible. \ourrep excels on charts requiring semantic coordination, such as mapping fields to grouped encodings, configuring sort orders, or creating multi-layer compositions. Ties or DirectVL wins occur on simpler charts (e.g., single-series bars, basic boxplots) that require minimal semantic reasoning. Results also show that an intermediate representation is especially valuable for models with limited capacity to manage free-form specifications. We show detailed experiment results and examples in \autoref{app:llm-eval-results}.

\section{Discussion and Conclusion}
\label{sec:discussion}

We introduced \ourrep, an intermediate language that uses a semantic type hierarchy to optimize chart configurations and generate library-native specifications. This abstraction opens new research opportunities:

\look{
\begin{itemize}[leftmargin=*, itemsep=2pt, parsep=0pt]
\item \emph{Extending to new visualization languages.} Emerging languages such as Gofish~\cite{pollock2025gofish} and PiCCL~\cite{shi2025piccl} introduce new visualization capabilities but often come with steep learning curves and limited training data. With \ourrep, researchers can implement a new backend that compiles \ourrep specifications to the target language. Downstream users can then plug the new language into existing tools without rewriting authoring interfaces, retraining agents, or learning a new low-level grammar from scratch (see \autoref{app:gallery}).
\item \emph{Visualization recommendation.} Existing engines like Draco~\cite{yang2023draco,moritz2018formalizing} and CompassQL~\cite{wongsuphasawat2016towards} rely on physical data representations, which introduce many rules over low-level parameters. With \ourrep's abstract semantic layer, recommenders can focus purely on high-level design choices (chart types and channel mappings) while delegating downstream adjustments to the compiler. This enables engines to learn generalizable visual idioms with significantly less training data.
\end{itemize}
}

\noindent Overall, \ourrep bridges visualization research and engineering practice through concise specifications and extensible backends.

\acknowledgments{
\look{We thank supports from the NSFC grants (No. 62502523, No. U2436209), the Beijing Natural Science Foundation (L247027), the Fundamental Research Funds for the Central Universities, and the Research Funds of Renmin University of China. }
}

\bibliographystyle{abbrv-doi-hyperref}

\bibliography{llmvis}

\newpage
\appendix
\section{Semantic type registry}

\noindent In \ourrep, we currently organize semantic types into 44 fine-grained L3 semantic types, grouped under 15 L2 families and 6 L1 domains. Our system maintains a semantic type registry that maps each semantic type to a set of abstract compilation properties that are chart- and encoding-agnostic, including parsing strategy, preferred encoding types, aggregation behavior, domain shape, zero-baseline semantics, and formatting defaults that influence chart configuration in later stages. We label each property according to the semantic level that primarily determines it (L1--L3). Properties labeled L2/L3 indicate that the property is generally introduced at the family level but may be further refined by individual semantic types within that family.

\begin{itemize}[leftmargin=*]\itemsep-1pt
\item \emph{Parsers (L1)}: Whether the raw data should be tested and parsed with temporal, numerical, or string parsers and how to process missing, null and NaN values.
\item \emph{Encoding Candidates (L2/L3)}: The set of semantically appropriate encoding types for a field, such as \textit{quantitative}, \textit{ordinal}, \textit{nominal}, or \textit{temporal}. This property defines valid casts from semantic data types to visual encoding types and prevents invalid assignments. For example, it allows a \textit{YearMonth} field to be cast to an ordinal encoding when necessary (e.g., to fit on the $y$-axis of a pyramid chart), while prohibiting the direct use of a \textit{DateTime} field as bar height.
\item \emph{Zero Baseline (L2)}: Specifies whether zero should be treated as a \textit{meaningful} baseline (e.g., Revenue), an \textit{arbitrary} one (e.g., Year), a \textit{contextual} one (e.g., Percentage), or \textit{none} when the notion of zero is irrelevant.
\item \emph{Aggregation Role (L2):} Specifies whether a field should function as a grouping variable, a record-level key, or an aggregation target, and if aggregated, which aggregation behaviors are semantically valid. We classify fields as \code{additive} (e.g., Revenue, Count), \code{intensive} (e.g., Percentage, Price), \code{signed-additive} (e.g., Profit), \code{dimension} for grouping fields such as Country, Status, and Month, or \code{identifier} for record-level keys such as OrderID and CustomerID.
\item \emph{Diverging Class (L2/L3)}: Whether a field semantically implies diverging treatment in visual encodings, such as diverging color scales or centered positional mappings, categorized as \code{inherent}, \code{conditional}, or \code{none}.
\item \emph{Domain Shape (L2/L3)}: Whether the scale implied by the data is \code{open}, \code{bounded}, \textit{fixed}, or \code{cyclic}, which in turn affects axis construction, tick generation, and domain padding.
\item \emph{Display Formatter (L3)}: Specifies how a field should be rendered in axis ticks and tooltips. In many cases, the preferred behavior is to preserve the raw representation when it is already readable; otherwise, the semantic type provides a more suitable formatter. For example, a \code{YearMonth} field may prefer a formatter such as \code{\%Y-\%m} if it is stored as timestamps, whereas a simple textual \code{Month} field may remain unchanged. This property allows the compiler to improve readability while preserving intended semantics.
\end{itemize}

\section{Semantic Type Error Resolution}
\label{app:semantic-error-resolution}
\begin{figure}[h]
    \centering
    \includegraphics[width=\linewidth]{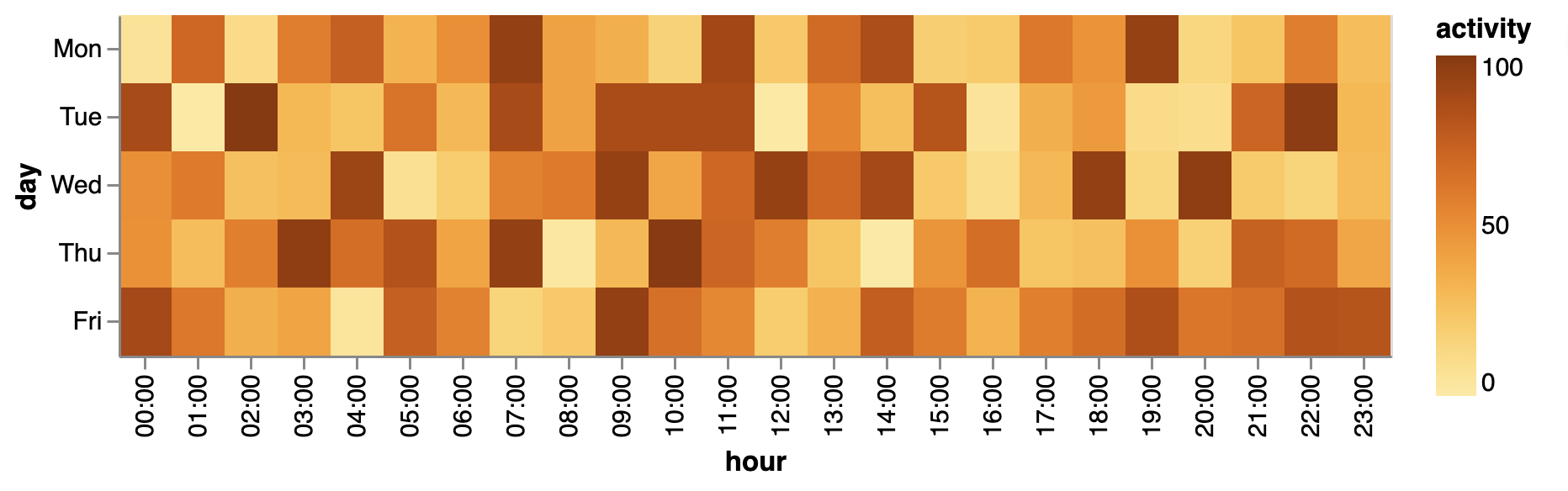}
    \vspace{-3mm}
    \caption{Day-of-week and hour-of-day are labeled as generic \code{Category} instead of the more precise \code{Day} and \code{Hour}.}
    \label{fig:semantic-type-fallback}
\end{figure}

In practice, semantic types are usually easy for both humans and AI agents to specify correctly, since they align with column names, value patterns, and common data semantics. When occasional mistakes occur, they are more often small ``near-misses'': for example, labeling a revenue field as \code{Count}, a month field as \code{Category}, or a temperature field as \code{Quantity}. \ourrep handles such cases gracefully, since its compiler frontend checks whether the declared semantic type is consistent with the field's raw data representation~\autoref{sec:compiler}. When the annotation is imprecise, it backs off along the type hierarchy to the most specific compatible ancestor, preserving as much semantic information as possible for downstream chart construction and optimization. If no compatible type can be found, the compiler ultimately falls back to the most generic type \code{Value}, relying primarily on the field's physical data type to perform best-effort inference and preserve chart quality. \autoref{fig:semantic-type-fallback} shows a heatmap example in which day-of-week and hour-of-day are annotated as generic \code{Category} rather than the more precise \code{Day} and \code{Hour}. Despite these imprecise annotations, the compiler still produces a correct and well-laid-out chart, but it loses some type-specific optimizations, such as treating the hour axis as a continuous temporal-like axis for better labeling.

\section{Gallery}
\label{app:gallery}
\begin{figure*}[htbp]
    \centering
    \includegraphics[width=0.995\linewidth]{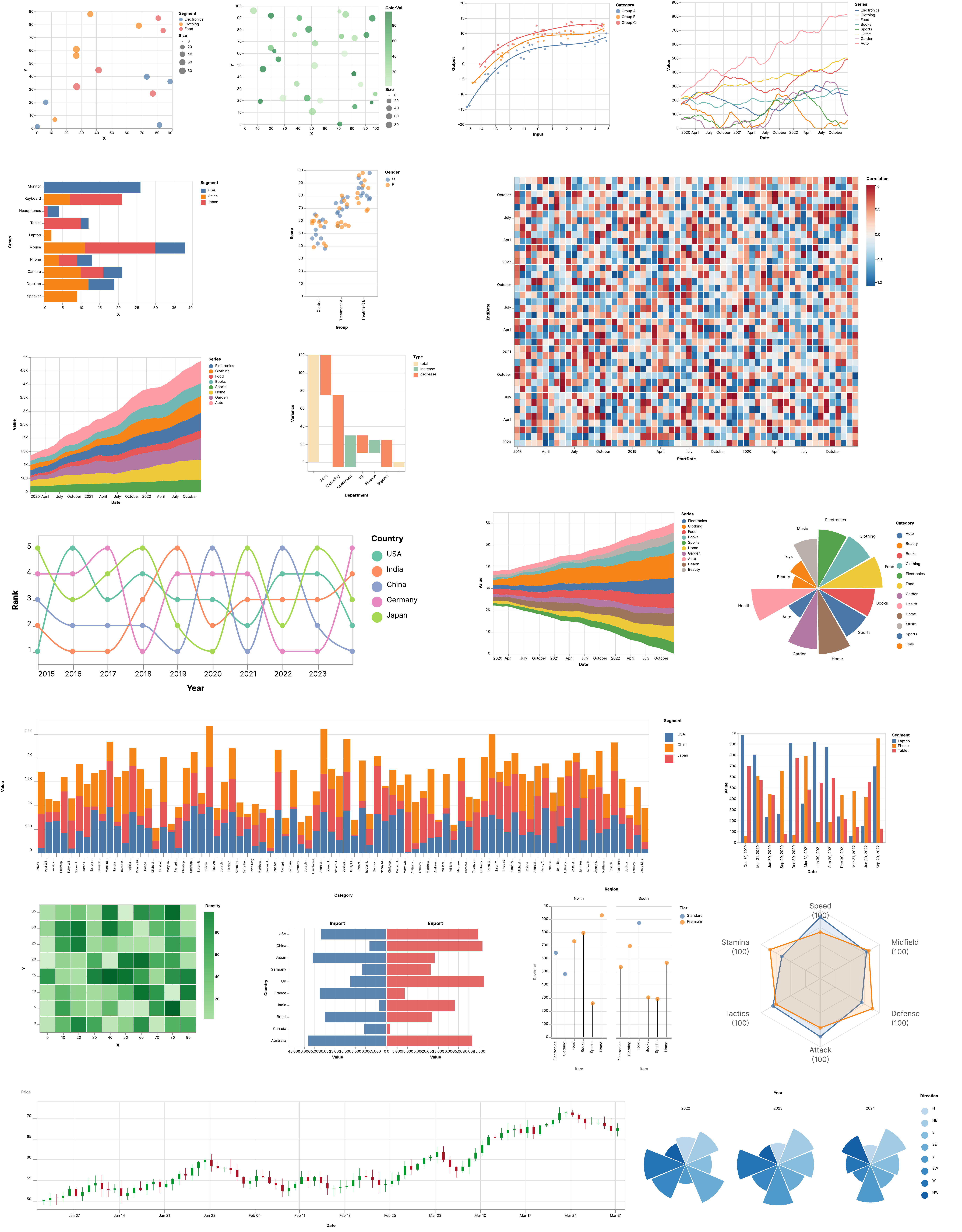}
    \caption{A gallery of standard visualization examples rendered using the Vega-Lite backend.}
    \label{fig:vegalite-examples}
\end{figure*}

\begin{figure*}[htbp]
    \centering
    \includegraphics[width=0.995\linewidth]{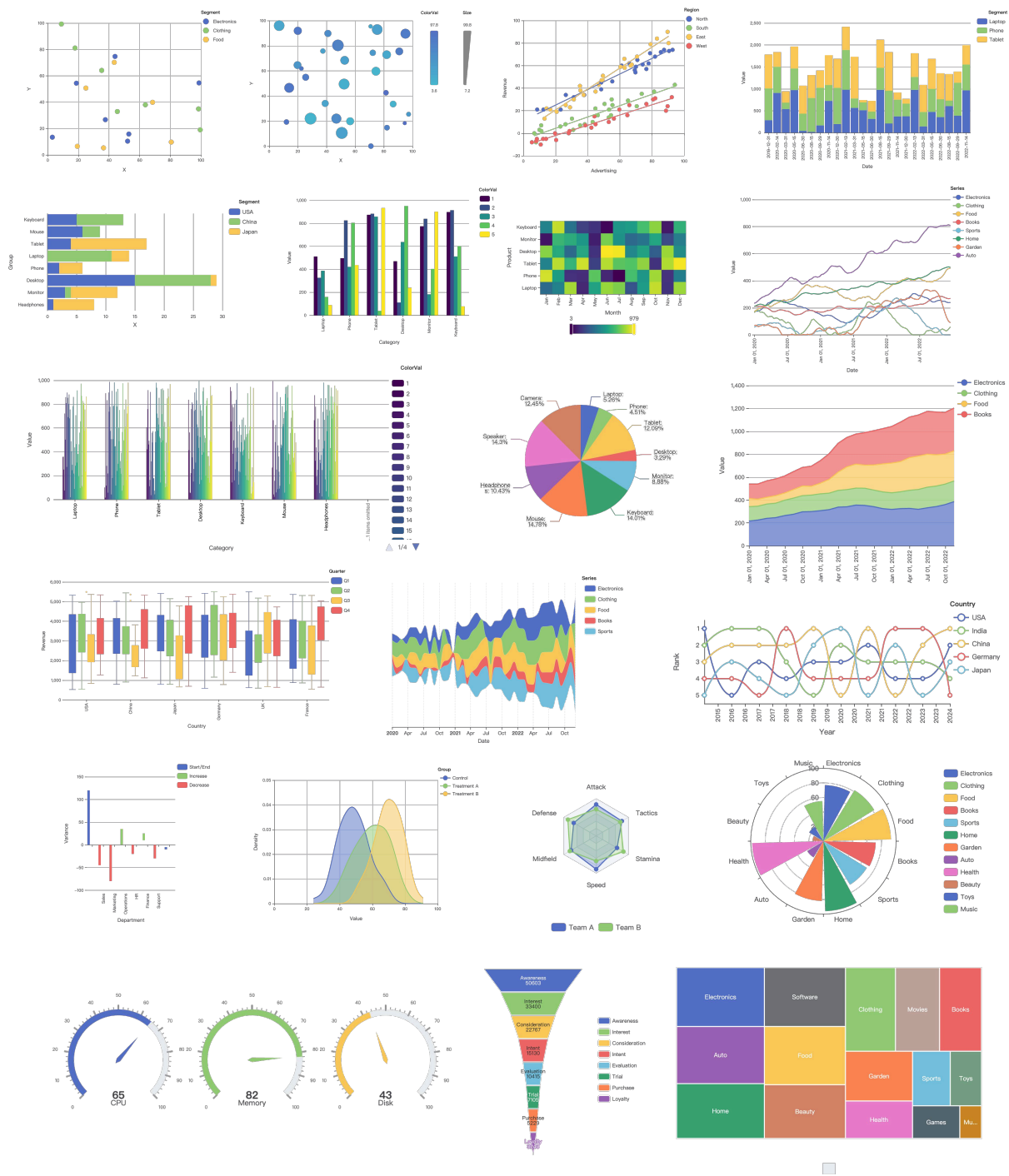}
    \caption{A selection of diverse chart types rendered using the ECharts backend.}
    \label{fig:echarts-examples}
\end{figure*}

\begin{figure*}[htbp]
    \centering
    \includegraphics[width=0.995\linewidth]{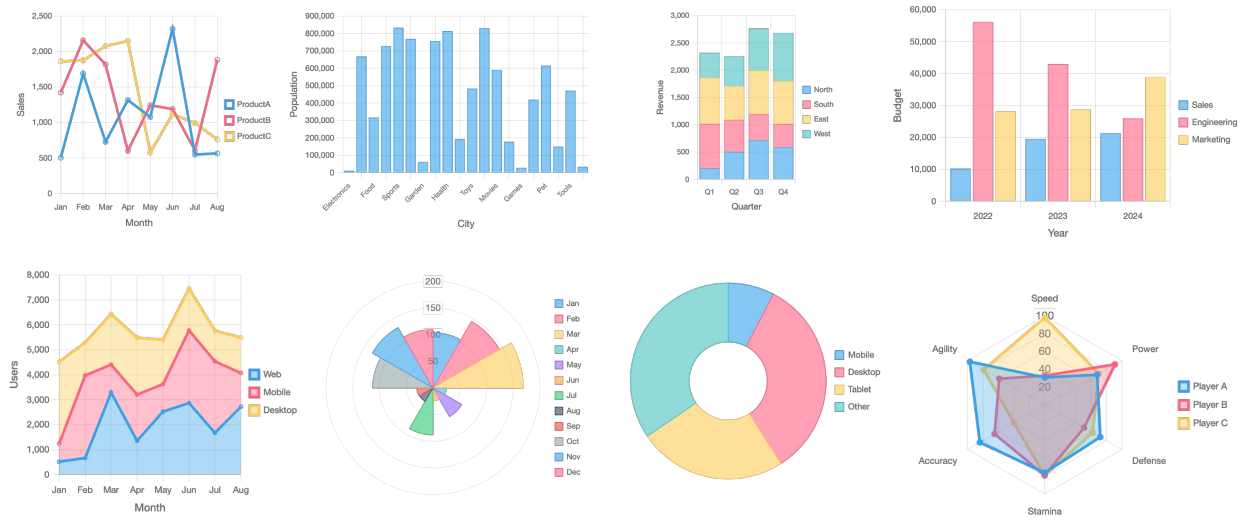}
    \caption{Chart.js renderings for common visualizations generated from \ourrep specifications.}
    \label{fig:chartjs-examples}
\end{figure*}

We present a gallery of visualization examples that demonstrate both the expressive power and the cross-backend portability of our proposed \ourrep language. So far, the gallery covers the following charts across backends; many of them are not natively supported by the libraries, and we implement them with dynamic templates. These charts automatically adapt to different data cardinalities and types thanks to \ourrep's semantic type-based optimization passes during the compilation process. \Cref{fig:vegalite-examples}--\Cref{fig:chartjs-examples} show example charts rendered in Vega-Lite, ECharts, and Chart.js. All visualizations are generated from the high-level \ourrep specifications, and the compiler determines the backend-specific low-level configurations based on semantic data types.

\begin{itemize}[leftmargin=*]
\item \textbf{Vega-Lite:} Scatter Plot, Linear Regression, Boxplot, Strip Plot, Bar Chart, Grouped Bar, Stacked Bar, Histogram, Lollipop, Pyramid, Line Chart, Dotted Line, Bump Chart, Area Chart, Streamgraph, Pie Chart, Rose Chart, Heatmap, Waterfall Chart, Density Plot, Ranged Dot Plot, Radar Chart, Candlestick Chart, US Map, and World Map.
\item \textbf{ECharts:} Scatter Plot, Boxplot, Bar Chart, Grouped Bar, Stacked Bar, Histogram, Heatmap, Line Chart, Area Chart, Streamgraph, Pie Chart, Candlestick Chart, Radar Chart, and Rose Chart.
\item \textbf{Chart.js:} Scatter Plot, Bar Chart, Grouped Bar, Stacked Bar, Histogram, Line Chart, Area Chart, Pie Chart, Radar Chart, and Rose Chart.
\end{itemize}

\begin{figure*}[htbp]
    \centering
    \includegraphics[width=0.995\linewidth]{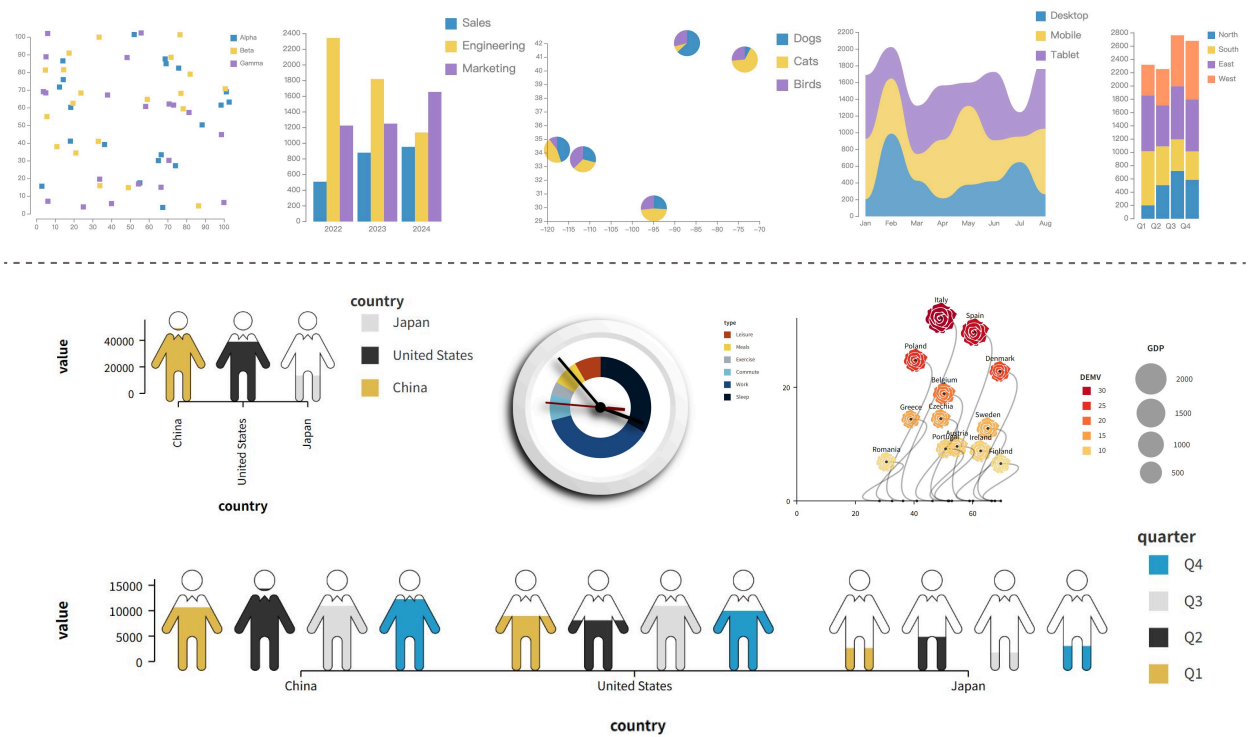}
    \caption{Extension experiments with Gofish (top) and PiCCL (bottom).}
    \label{fig:other-examples}
\end{figure*}

\bpstart{Extension Experiment} To illustrate the extensibility of our approach beyond traditional web-based libraries, \Cref{fig:other-examples} presents additional examples rendered via Gofish~\cite{pollock2025gofish} and PiCCL~\cite{shi2025piccl}. Because PiCCL is an infographics-oriented library, our extension currently relies on a fixed set of media assets and composition methods (overlay, replacement, fill) for each template. In future work, it would be worth exploring how data semantics could automatically guide the recommendation and generation of composition materials based on chart intent.

\newpage
\section{LLM evaluation setup details}
\label{app:llm-eval-setup}

In this section, we present details of our agent and study design. We use the {TidyTuesday} 2025 datasets~\footnote{\url{https://github.com/rfordatascience/tidytuesday}}, which include 63 tables spanning 51 real-world topics, with substantially larger scale (median 952 rows, 49\% exceeding 1000 rows, and median 8 columns). As this corpus lacks annotated queries, we employ an LLM agent to generate fine-grained exploration questions~\cite{DBLP:journals/corr/abs-2509-21825}, which serve as input prompts for visualization agents, enabling controlled diversity in both questions and chart types. Overall, our evaluation leverages the following LLM agents:
    \begin{itemize}[leftmargin=*]\itemsep-1pt
    \item \emph{Question generation agent}: given an input dataset, generates diverse visualization-oriented data analysis questions. In particular, we breakdown the score for each chart into three categories.
    \item \emph{Chart agents}: given the dataset and a question as inputs, generate a visualization. We use the \emph{Flint agent} introduced earlier, with Vega-Lite as the backend library. We further compare it with a DirectVL agent, which is instructed to generate a custom Vega-Lite specification that can answer the question.
    \item \emph{Grader agent}: given a generated chart specification, the corresponding question, and the input data, the grader assesses whether the output is correct, interpretable, and appropriate for the task based on the rendered image from the visualization.
    \end{itemize}
    Here, we use gpt-5.1 for question generation and grading, and we ablate chart agents with both gpt-5.1 and gpt-4.1 to compare the models' effect on chart generation quality.

\subsection{Question generation agent}

The question-generation agent lets us systematically construct a diverse set of visualization tasks for evaluation. The prompt encourages coverage of trends, comparisons, distributions, relationships, rankings, compositions, correlations, geographic patterns, and forecasting, and it asks the agent to include a suitable chart-type hint for each question from a fixed set of supported visualizations.

\begin{minted}[fontsize=\small,breaklines]{text}
You are a data analyst who generates questions about datasets for visualization and analysis.

Given a dataset summary, generate a list of questions that ask for visualizations to show trends/insights about the data.
Note that each question should be suitable for a single layered chart.

Generate questions of varying styles:
1. **Ambiguous/High-level**: Open-ended questions that allow for interpretation (e.g., "What are the key trends in this data?")
2. **Concrete/Low-level**: Specific questions about particular aspects (e.g., "How does sales vary by region?")
3. **Chart-specific**: Questions that directly request a specific chart type (e.g., "Show me a bar chart of revenue by category").
    Do not ask for multi-layered or annotated charts. Chart types can be selected from "Scatter Plot", "Bar Chart", "Line Chart", "Area Chart", "Histogram", "Heatmap", "Group Bar Chart", "Boxplot".

Output a JSON array of questions with the following format:
```json
[
    {
        "question": "...",
        "style": "ambiguous" | "concrete" | "chart_specific",
        "complexity": "low" | "medium" | "high",
    },
    ...
]
```

Within the number of questions requested by the user, generate a diverse set of questions that cover different aspects of the data.
\end{minted}

\subsection{Chart agent (\ourrep)}

This section presents the prompt used for the \ourrep chart agent. Although \ourrep is a new intermediate language, its simple syntax makes it easy for LLM agents to use. A prompt sketch is shown below. Here, we specifically instruct the agent to follow the semantic hierarchy so it can generate \texttt{dataSpec} based on data context information. Once \texttt{dataSpec} is generated, it can be reused by the user when adapting the chart. This improves token efficiency, reduces latency, and makes iterative visualization workflows faster and more stable.

\begin{tcolorbox}[
  colback=gray!5,
  colframe=gray!50,
  boxrule=0.3pt,
  arc=2pt,
  left=4pt,right=4pt,top=4pt,bottom=4pt
]
\begin{minted}[fontsize=\small,breaklines]{text}
[SYSTEM]
- Semantic type hierarchy guidance for fields
- Visualization design space:
  chart type | available channels | chart-specific properties
- guidance for SQL/Python preprocessing scripts
- guidance for choosing chart types

[OUTPUT FORMAT]
- Output schema guidance for dataSpec and chartSpec
- A python script that transforms the data if needed

[CONTEXT]
- Input tables, columns, file paths
- Field summaries
- Sample rows

[QUESTION]
- User's natural-language request or partial specification
\end{minted}
\end{tcolorbox}

\begin{minted}[fontsize=\small,breaklines]{text}
You are a data scientist who recommends data and visualizations.
Given [CONTEXT] (dataset summaries) and [GOAL] (user intent), recommend a transformed dataset and visualization, then write a Python script to produce it.

**[SEMANTIC TYPE REFERENCE]**

Choose the most specific type that fits. Only annotate fields used in chart encodings.

| Category | Types |
|---|---|
| Temporal | DateTime, Date, Time, Timestamp, Year, Quarter, Month, Week, Day, Hour, YearMonth, YearQuarter, YearWeek, Decade, Duration |
| Monetary measures | Amount, Price |
| Physical measures | Quantity, Temperature |
| Proportion | Percentage |
| Signed/diverging | Profit, PercentageChange, Sentiment, Correlation |
| Generic measures | Count, Number |
| Discrete numeric | Rank, Score |
| Identifier | ID |
| Geographic | Latitude, Longitude, Country, State, City, Region, Address, ZipCode |
| Entity names | Category, Name |
| Coded categorical | Status, Boolean, Direction |
| Binned ranges | Range |
| Fallback | Unknown |

**[FIELD METADATA REFERENCE]**

For each field used in chart encodings, provide:
- "semantic_type": one type from [SEMANTIC TYPE REFERENCE]
- "intrinsic_domain": [min, max] when the field has a known bounded scale
    - Infer from data values and context: e.g., if a "rating" column has values 1-10, the domain is [1, 10]; if it is clearly a 5-star system, use [1, 5].
    - For Percentage: [0, 100] if values are whole-number percentages, [0, 1] if fractional.
    - For Correlation: always [-1, 1].
    - Do NOT provide for open-ended measures like Amount, Count, Quantity, Temperature, etc.
    - Only provide when the scale bounds are clear from the data or domain knowledge.
- "mid": a meaningful midpoint or neutral reference value
    - Use for signed/diverging fields when relevant.
    - Examples: 0 for Profit, PercentageChange, Sentiment, Correlation; 32 for Fahrenheit temperature only when the freezing point is semantically meaningful to the task.
    - Do NOT provide unless there is a clear semantic midpoint.
- "unit": a short unit string for physical/monetary quantities
- "sort_order": the natural order for ordinal string fields
    - Provide only if the field is string type and ordinal.
    - Examples: month names, weekdays, ranked categories, ranges, education levels.
    - Do NOT provide when alphabetical order is already correct or there is no inherent order.
    - Examples where sort_order is usually unnecessary: Name, State, City.

Examples:
- Use **Amount** for summed monetary totals, **Price** for per-unit prices, **Profit** for values that can be negative.
- Use **Temperature** (not Quantity) for temperature.
- Use **Year** (not Number) for columns like "year" with values 2020, 2021.

**[CHART TYPE REFERENCE]**

| chart_type | encodings | config |
|---|---|---|
| Scatter Plot | x, y, color, size, facet | opacity (0.1–1.0) |
| Regression | x, y, color, size, facet | regressionMethod ("linear","log","exp","pow","quad","poly"), polyOrder (2–10) |
| Bar Chart | x, y, color, facet | — |
| Grouped Bar Chart | x, y, group, facet | — |
| Line Chart | x, y, color, strokeDash, facet | interpolate ("linear","monotone","step") |
| Area Chart | x, y, color, facet | — |
| Heatmap | x, y, color, facet | colorScheme ("viridis","blues","reds","oranges","greens","blueorange","redblue") |
| Boxplot | x, y, color, facet | — |
| Pie Chart | size, color, facet | innerRadius (0–100; 0=pie, >0=donut) |
| Lollipop Chart | x, y, color, facet | — |
| Waterfall Chart | x, y, color, facet | — |
| Candlestick Chart | x, open, high, low, close, facet | — |
| World Map | longitude, latitude, color, size | projection ("mercator","equalEarth","naturalEarth1","orthographic"), projectionCenter ([lon,lat]) |
| US Map | longitude, latitude, color, size | — (fixed albersUsa) |

**Critical chart rules:**
- **facet**: available for all chart types; use a categorical field with small cardinality.
- All fields in "encodings" must also appear in "output_fields". Typically use 2–3 channels (x, y, color/size).

You will produce two outputs: a JSON spec (```json```) and a Python script (```python```). No extra text.

**Step 1: JSON spec** — infer user intent and recommend a visualization.

```json
{
    "display_instruction": "", // short verb phrase (<12 words) capturing computation intent. Bold **column names** (semantic matches count). For follow-ups, describe only the new part.
    "input_tables": [...],     // table names from [CONTEXT] to use. Table 1 is the currently viewed table — prioritize it.
    "output_fields": [...],    // desired output fields (include intermediate fields)
    "chart": {
        "chart_type": "",      // from [CHART TYPE REFERENCE]
        "encodings": {},       // visual channels -> output field names
        "config": {}           // optional styling
    },
    "field_metadata": {        // metadata for each encoding field only
        "<field>": {
            "semantic_type": "Category",
            "intrinsic_domain": [0, 100], // optional
            "mid": 0,                     // optional
            "unit": "%",                  // optional
            "sort_order": [...]           // optional
        }
    },
    "output_variable": ""      // descriptive snake_case name (e.g. "sales_by_region"), not "result_df"
}
```

**field_metadata rules:**
- Include one entry for every field used in chart encodings.
- "semantic_type" is required.
- "intrinsic_domain", "mid", "unit", and "sort_order" are optional and should be included only when clearly justified.
- Do not fabricate units, domains, or ordered categories if they are not supported by the data or context.

**Data format rules:**
- Output must be tidy (one field per visual channel, like VegaLite/ggplot2).
- For multiple similar columns: reshape to long format (only same semantic type in one column).
- For derived metrics: compute new fields (correlation, difference, profit, etc.).
- Keep encodings to 2–3 channels (x, y, color/size). Add facet only when needed.

**Step 2: Python script** — transform input data to produce a DataFrame with all "output_fields". Keep it simple and readable. The script MUST assign the final result to the variable named in "output_variable" from Step 1.
\end{minted}

\subsection{Chart agent (Vega-Lite)}

This section provides the prompt we use for the Vega-Lite visualization agent used in our experiment. Here, the agent is instructed to directly generate a Vega-Lite spec for visualization. We instructed the agent not to generate custom annotations to avoid diverting the grader's attention to annotations that could interfere with the comparison.

\begin{minted}[fontsize=\small,breaklines]{text}
You are a data scientist helping users create visualizations.
The user will provide you information about what visualization they would like to create, and your job is to:
1. Transform the data using a standalone Python script
2. Generate VegaLite visualization code directly

The recommendation and transformation should be based on the [CONTEXT] and [GOAL] provided by the user.
The [CONTEXT] shows what the current dataset is, and the [GOAL] describes what visualization the user wants.

Your output should consist of three parts:

1. A JSON object with metadata about the visualization:
```json
{
    "recap": "...",
    "display_instruction": "...",
    "recommendation": "...",
    "output_fields": [...],
    "output_variable": "..."
}
```
- "output_variable": descriptive snake_case name (e.g. "sales_by_region"), not "result_df"

2. A standalone Python script that reads data from files and produces a DataFrame:

```python
import pandas as pd
import numpy as np

# Read data from workspace files (use exact filenames from [CONTEXT])
df = pd.read_parquet('table_name.parquet')

# Transform as needed
# ...

# Assign result to the variable named in "output_variable"
output_variable_name = transformed_df
```

Note:
- The script is standalone (no function wrapper needed)
- Read data directly from files using the paths shown in [CONTEXT]
- The final result DataFrame MUST be assigned to the variable named in "output_variable" from the JSON spec
- datetime handling: convert years to numbers, year-month/year-month-day to strings
- Never return datetime objects directly

3. VegaLite visualization code in JSON format:

```json
{
  "$schema": "https://vega.github.io/schema/vega-lite/v5.json",
  "data": {"name": "data"},
  "mark": "...",
  "encoding": {
    "x": {"field": "...", "type": "..."},
    "y": {"field": "...", "type": "..."},
    ...
  },
  ...
}
```

Important notes for VegaLite code:
- Always use `"data": {"name": "data"}` - the transformed dataframe will be provided with this name
- Field types should be one of: "quantitative", "nominal", "ordinal", "temporal"
- You can use any VegaLite mark type: circle, bar, line, area, rect, point, etc.
- You may use "layer" when a chart type requires multiple marks (e.g., lollipop = rule + circle). Do NOT use layers for annotations, text overlays, or combining separate charts.
- Do NOT add a chart "title", "subtitle", or "description" property
- Do NOT add tooltip encoding or config.mark.tooltip
- Do NOT add custom axis titles or legend titles — let VegaLite use field names as defaults
- Do NOT use "transform" in VegaLite — all data transformations must be done in Python
- Do NOT add annotations, text marks, or selection/interaction parameters
- Keep styling minimal: only specify mark, encoding, width, and height

The output must contain:
1. One JSON object with metadata
2. One Python code block (standalone script)
3. One JSON code block with VegaLite spec

Do not add any extra text explanation outside these three blocks.
\end{minted}

\subsection{Grading agent}

The grading agent uses a checklist-based evaluation protocol to assess each agent response. It is implemented as a vision-language model that receives the rendered chart image and judges the quality of the visualization directly. The four evaluation dimensions capture whether the chart answers the user’s question, whether it contains errors, and how clear and well-designed it is.

\begin{minted}[fontsize=\small,breaklines]{text}
You are an expert evaluator for data visualization and analysis tasks.

You will be given:
1. **Data Summary**: Information about the input dataset
2. **Task Description**: What the user asked for
3. **Chart Image**: The generated visualization
4. **Chart Caption**: A short text explaining what the chart shows
5. **Rubric**: Evaluation criteria with scoring guidelines

Evaluate the chart based on the rubric and provide:
1. A score for each rubric criterion
2. Brief justification for each score
3. An overall score

Output your evaluation as JSON:
```json
{
    "criteria_scores": {
        "<criterion_name>": {
            "score": <number>,
            "max_score": <number>,
            "justification": "<brief explanation>"
        },
        ...
    },
    "overall_score": <number>,
    "overall_max_score": <number>,
    "summary": "<brief overall assessment>"
}

Evaluation Criteria (each scored 0-5):

1. **Relevance** (0-5): Does the visualization address the task/question?
   - 5: Perfectly addresses the task
   - 3: Partially addresses the task
   - 0: Does not address the task at all

2. **Chart Errors** (0-5): Are there any errors in the chart? E.g., skewed axis, missing data, unreadable content, etc.
   - 5: No errors in the chart
   - 3: Minor errors in the chart
   - 0: Major errors in the chart

3. **Clarity** (0-5): Is the visualization clear and easy to understand?
   - 5: Very clear representation of the data that is easy to understand.
   - 3: Somewhat clear but could be improved (e.g., requires some efforts to read the patterns and details to answer the question)
   - 0: Confusing, poorly labeled, or misleading, difficult to understand the data or the trend.

4. **Design Quality** (0-5): Is the visualization well-designed and visually effective?
   - 5: Excellent use of colors, scales, and visual encoding, clear and effective.
   - 3: Acceptable design with some suboptimal choices, could be improved.
   - 0: Poor design choices that hinder understanding.
```
\end{minted}

\section{LLM Evaluation results}
\label{app:llm-eval-results}

\bpstart{Scores} We summarize the quantitative evaluation results of \ourrep and DirectVL in the following tables. They report overall performance across models, per-criterion breakdowns, and finer-grained analyses by chart type, providing a compact view of the evaluation from aggregate results to detailed category-level performance.
\Cref{tab:eval-criteria-breakdown}--\Cref{tab:eval-chart-type} summarize the quantitative evaluation of \ourrep and DirectVL. Together, they present overall results across models, per-criterion breakdowns, and chart-type-specific analyses. These evaluation scores are used for pairwise comparison in our evaluation.

\begin{table*}[t]
\centering
\caption{Breakdown by grading criterion for each agent-model pair (mean scores; Overall is out of 20, other criteria are out of 5).}
\label{tab:eval-criteria-breakdown}
\begin{tabular}{llrrrrr}
\toprule
Model & Agent & Overall & Relevance & Chart Errors & Clarity & Design Quality \\
\midrule
\multirow{2}{*}{gpt-5.1} & \ourrep & \textbf{16.27 $\pm$ 3.37} & \textbf{4.60} & \textbf{4.11} & \textbf{3.87} & \textbf{3.68} \\
 & DirectVL & 15.91 $\pm$ 3.53 & 4.45 & 4.04 & 3.80 & 3.55 \\
\midrule
\multirow{2}{*}{gpt-5-mini} & \ourrep & \textbf{16.16 $\pm$ 3.56} & \textbf{4.51} & \textbf{4.06} & \textbf{3.91} & \textbf{3.68} \\
 & DirectVL & 15.60 $\pm$ 3.42 & 4.39 & 3.90 & 3.74 & 3.58 \\
\midrule
\multirow{2}{*}{gpt-4.1} & \ourrep & \textbf{15.91 $\pm$ 3.46} & \textbf{4.43} & \textbf{4.12} & \textbf{3.79} & \textbf{3.57} \\
 & DirectVL & 15.34 $\pm$ 3.87 & 4.26 & 3.90 & 3.67 & 3.48 \\
\bottomrule
\end{tabular}%
\end{table*}

\begin{table*}[h]
\centering
\caption{Chart-type breakdown of overall GPT-5.1 scores (mean over questions, out of 20).}
\label{tab:eval-chart-type}
\begin{tabular}{lrrr}
\toprule
Chart Type & $n$ & \ourrep & DirectVL \\
\midrule
Area Chart & 3 & \textbf{18.67} & 17.00 \\
Bar Chart & 59 & 15.68 & \textbf{15.69} \\
Boxplot & 43 & 15.93 & \textbf{16.74} \\
Grouped Bar Chart & 46 & \textbf{16.37} & 15.74 \\
Heatmap & 22 & \textbf{16.41} & 15.59 \\
Line Chart & 45 & \textbf{16.69} & 15.75 \\
Lollipop Chart & 49 & \textbf{16.06} & 15.28 \\
Pie Chart & 8 & 14.62 & \textbf{14.75} \\
Scatter Plot & 36 & \textbf{17.67} & 17.36 \\
US Map & 1 & \textbf{17.00} & 14.00 \\
World Map & 3 & \textbf{12.00} & 10.67 \\
\bottomrule
\end{tabular}
\end{table*}

\bpstart{LLM-generated examples and charts} We next show example questions, responses, and grading results for LLM charts generated in this experiment.

\newcommand{\examplescale}{0.3}

\begin{figure}[ht]
    \centering
    \includegraphics[scale=\examplescale]{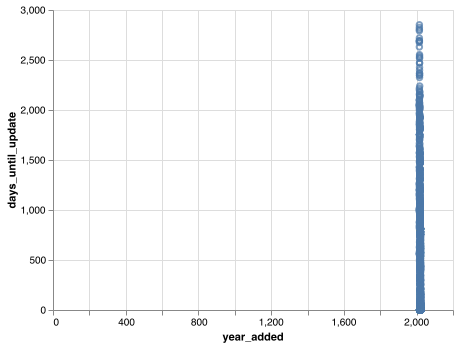}

    \vspace{1mm}
    {\footnotesize DirectVL}

    \vspace{2mm}
    \includegraphics[scale=\examplescale]{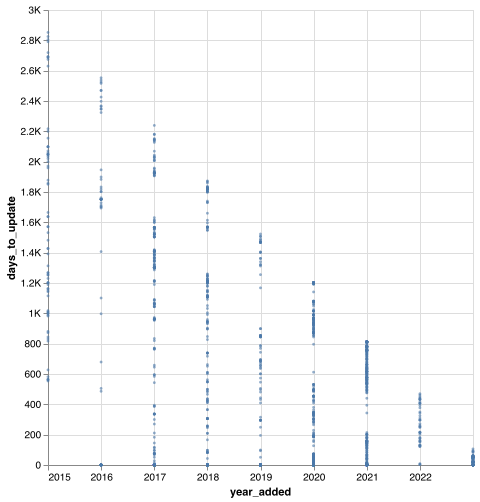}

    \vspace{1mm}
    {\footnotesize Flint}

    \caption{Example 1 on \texttt{apisguru\_apis}. Top: DirectVL. Bottom: \ourrep.}
    \label{fig:appendix-ex1}
\end{figure}

\paragraph{Example 1 (\ourrep is better).}
In this example, the agent is asked to generate a scatter plot for the \texttt{apisguru\_apis} dataset. The resulting charts are shown in \Cref{fig:appendix-ex1}. Here, the grader recognizes that the visualization generated by the DirectVL agent has $x$-axis rendering issues causing skewed charts, though it is gentle in discounting its score since the encoding is right. The chart from \ourrep received a full score.

\begin{minted}[breaklines]{text}
dataset:
  apisguru_apis

question:
  Use a scatter plot to show the relationship between the year an API spec was first added and the total number of days until it was last updated (difference between updated and added), with each point representing one row.

scores:
  flint: 20
  directvl: 16

grader_reasoning:
  flint: "A clear, accurate, and well-designed scatter plot that fully satisfies the specified task."
  directvl: "The scatter plot correctly addresses the task and is error-free, but the wide x-axis range compresses the data and limits clarity and visual effectiveness."
\end{minted}

\begin{figure}[ht]
    \centering
    \includegraphics[scale=\examplescale]{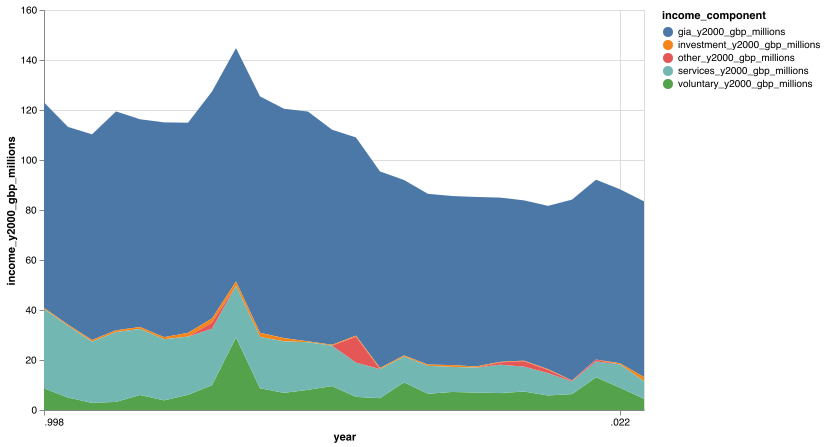}

    \vspace{1mm}
    {\footnotesize DirectVL}

    \vspace{2mm}
    \includegraphics[scale=\examplescale]{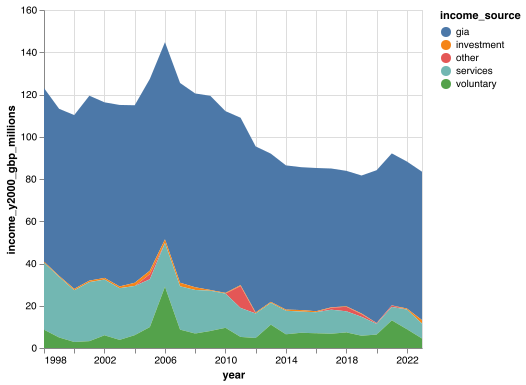}

    \vspace{1mm}
    {\footnotesize Flint}

    \caption{Example 2 on \texttt{bl\_funding}. Top: DirectVL. Bottom: \ourrep.}
    \label{fig:appendix-ex2}
\end{figure}

\paragraph{Example 2 (\ourrep is better).}
In this example, the agent is asked to generate a stacked area chart for the \texttt{bl\_funding} dataset. The resulting charts are shown in \Cref{fig:appendix-ex2}. The DirectVL agent in fact parsed the data incorrectly and thus received a lower score.

\begin{minted}[breaklines]{text}
dataset:
  bl_funding

question:
  Use a stacked area chart to show the composition of year_2000-adjusted income (gia_y2000_gbp_millions, voluntary_y2000_gbp_millions, investment_y2000_gbp_millions, services_y2000_gbp_millions, other_y2000_gbp_millions) over time by year.

scores:
  flint: 20
  directvl: 17

grader_reasoning:
  flint: "Well-executed stacked area chart that correctly and clearly shows the composition of year_2000-adjusted income by source over time, with no detectable errors and strong visual design."
  directvl: "A relevant and generally well-executed stacked area chart that correctly shows the composition of year-2000-adjusted income over time, with minor issues in x-axis labeling and layout affecting clarity and design."
\end{minted}

\begin{figure}[ht]
    \centering
    \includegraphics[scale=\examplescale]{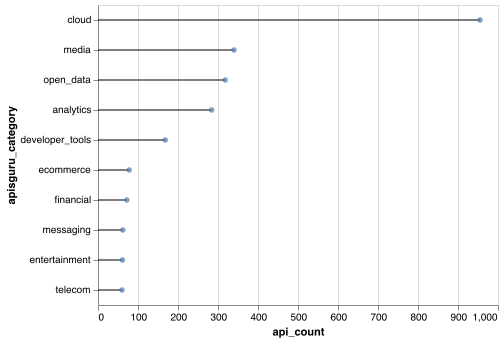}

    \vspace{1mm}
    {\footnotesize DirectVL}

    \vspace{2mm}
    \includegraphics[scale=\examplescale]{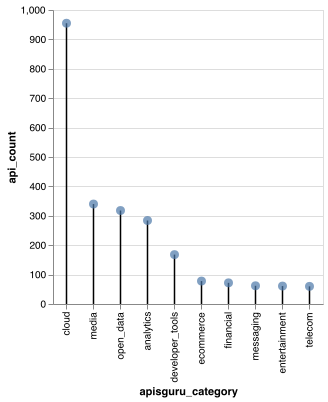}

    \vspace{1mm}
    {\footnotesize Flint}

    \caption{Example 3 on \texttt{api\_categories}. Top: DirectVL. Bottom: \ourrep.}
    \label{fig:appendix-ex3}
\end{figure}

\paragraph{Example 3 (\ourrep is worse).}
In this example, the agent is asked to generate a lollipop chart for the \texttt{api\_categories} dataset. The resulting charts are shown in \Cref{fig:appendix-ex3}. \ourrep received a slightly lower score because the horizontal lollipop generated by the DirectVL agent had a better layout.

\begin{minted}[breaklines]{text}
dataset:
  api_categories

question:
  Use a lollipop chart to rank apisguru_category values by their API counts and highlight the top 10 most common categories.

scores:
  flint: 18
  directvl: 19

grader_reasoning:
  flint: "Accurate and effective lollipop chart that fulfills the task by clearly presenting the top 10 API categories by count, with minor opportunities to improve label readability and visual emphasis."
  directvl: "A clear and accurate lollipop chart that correctly ranks and displays the top 10 API categories by count, with solid design and no visible errors."
\end{minted}

\begin{figure}[ht]
    \centering
    \includegraphics[scale=0.255]{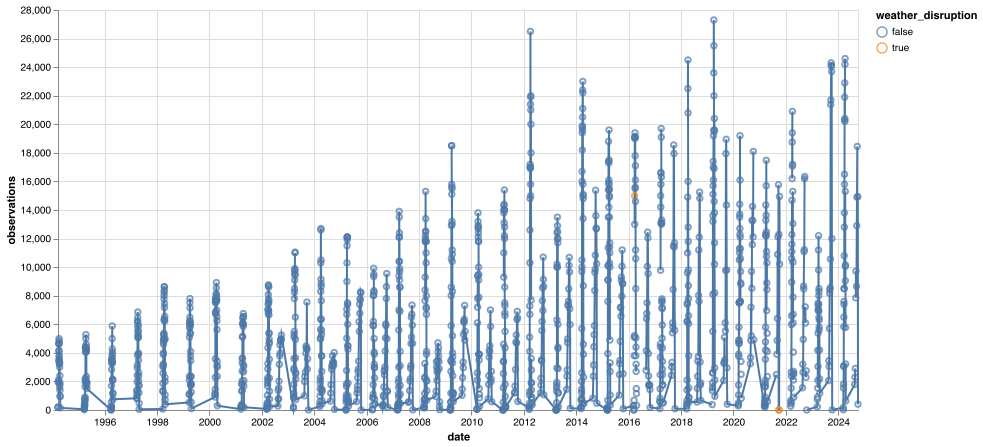}

    \vspace{1mm}
    {\footnotesize DirectVL}

    \vspace{2mm}
    \includegraphics[scale=\examplescale]{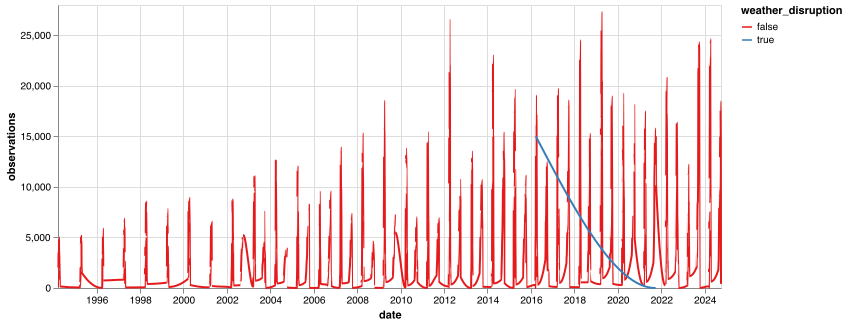}

    \vspace{1mm}
    {\footnotesize Flint}

    \caption{Example 4 on \texttt{cranes}. Top: DirectVL. Bottom: \ourrep.}
    \label{fig:appendix-ex4}
\end{figure}

\paragraph{Example 4 (\ourrep is worse).}
In this example, the agent is asked to generate a line chart for the \texttt{cranes} dataset. The resulting charts are shown in \Cref{fig:appendix-ex4}. This is a quite interesting example, \ourrep decides that it's two line series and decides to connect two disruption points, which in fact should not be connected, thus received lower scores.

\begin{minted}[breaklines]{text}
dataset:
  cranes

question:
  Plot a line chart of daily crane observations over time, coloring points by whether there was a weather disruption, to see long-term trends and how disruptions align with peaks and troughs.

scores:
  flint: 11
  directvl: 18

grader_reasoning:
  flint: "The chart generally encodes the right variables and shows long-term trends, but likely uses the wrong or additional data, relies on continuous lines for a boolean condition, and is not optimized for clearly seeing how weather disruptions align with peaks and troughs."
  directvl: "A relevant and technically sound visualization that effectively shows long-term crane observation trends and highlights weather disruptions, though the dense plotting and continuous lines make it somewhat busy."
\end{minted}

\begin{figure}[t]
    \centering
    \includegraphics[scale=\examplescale]{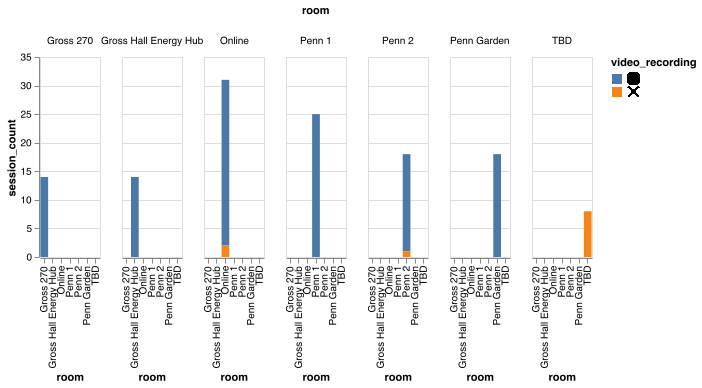}

    \vspace{1mm}
    {\footnotesize DirectVL}

    \vspace{2mm}
    \includegraphics[scale=\examplescale]{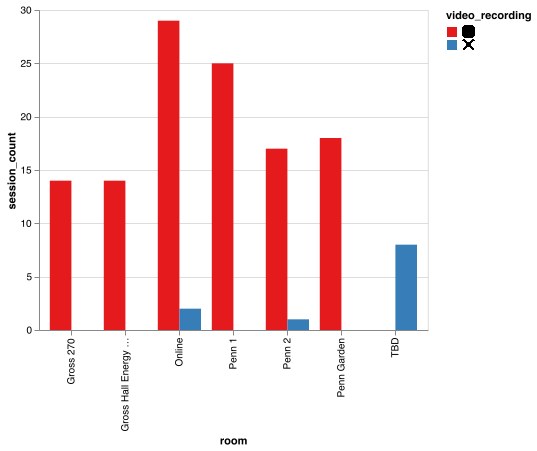}

    \vspace{1mm}
    {\footnotesize Flint}

    \caption{Example 5 on \texttt{user2025}. Top: DirectVL. Bottom: \ourrep.}
    \label{fig:appendix-ex5}
\end{figure}

\paragraph{Example 5 (Grader mistake).}
\autoref{fig:appendix-ex5} shows a case where the grader appears to misjudge the outputs, assigning \ourrep a score of 10 versus 13 for DirectVL. The underlying dataset is skewed for this question, which seems to have led the grader to penalize both charts. However, \ourrep actually produces the faithful grouped bar chart requested in the prompt, whereas DirectVL produces an incorrect faceted stacked bar chart instead. Although the grader can be imprecise in individual cases like this, the aggregate comparison over 315 runs still provides a useful signal for understanding overall agent behavior.

\begin{minted}[breaklines]{text}
dataset:
  user2025

question:
  Create a grouped bar chart comparing the number of sessions with and without video recordings (video_recording) for each room (room).

scores:
  flint: 10
  directvl: 13

grader_reasoning:
  flint: "The chart captures the correct variables but fails to produce a true grouped bar chart: bars for recording status overlap per room, obscuring data and limiting comparability. With side-by-side grouping and clearer labels, it would more effectively fulfill the task."
  directvl: "The visualization encodes the correct counts of sessions with and without video recordings by room, but the choice to facet by room while also using room on the x-axis results in redundancy and reduced clarity. A single grouped bar chart with rooms along one axis would better satisfy the task and improve readability and design."
\end{minted}

\end{document}